\documentclass[aps, twocolumn, epsf, floats, pre, nofootinbib]{revtex4-1}
\usepackage{graphics,graphicx,epsfig}
\usepackage{amssymb,color}
\usepackage{epsf,epstopdf,wrapfig}
\usepackage{amsmath}
\usepackage{amsmath}
\usepackage{amssymb}
\usepackage{bbold}
\usepackage{graphicx}
\usepackage{bm}
\usepackage{xr}

\def\(({\left(}
\def\)){\right)}                       
\def\[[{\left[}
\def\]]{\right]}

\newcommand{\beq}{\begin{equation}}
\newcommand{\eeq}{\end{equation}}
\newcommand{\bea}{\begin{eqnarray}}
\newcommand{\eea}{\end{eqnarray}}

\begin{document}

\title{Spatio-temporal correlations in models of collective motion ruled by different \\ dynamical laws}

\author{
Andrea Cavagna$^{1}$, 
Daniele Conti$^{2}$,
Irene  Giardina$^{1,2}$,
Tomas S. Grigera$^{3,4}$
Stefania Melillo$^{1,2}$, 
Massimiliano Viale$^{1,2}$
}

\affiliation{$^1$ Istituto Sistemi Complessi, Consiglio Nazionale delle Ricerche, UOS Sapienza, 00185 Rome, Italy}
\affiliation{$^2$ Dipartimento di Fisica, Universit\`a\ Sapienza, 00185 Rome, Italy}
\affiliation{$^3$ Instituto de F\'\i{}sica de L\'\i{}quidos y Sistemas Biol\'ogicos
  (IFLYSIB), CONICET ---  Universidad Nacional de La Plata, Calle 59 no.~789, B1900BTE La
  Plata, Argentina}
\affiliation{$^4$ CCT CONICET La Plata, Consejo Nacional de Investigaciones
  Cient\'\i{}ficas y T\'ecnicas, Argentina}

\begin{abstract}

Information transfer is an essential factor in determining the robustness of biological systems with distributed control. The most direct way to study the mechanisms ruling  information transfer is to experimentally observe the propagation across the system of a signal triggered by some perturbation. However, this method may be inefficient for experiments in the field, as the possibilities to perturb the system are limited and empirical observations must rely on natural events. An alternative approach is to use spatio-temporal correlations to probe the information transfer mechanism directly from the spontaneous fluctuations of the system, without the need to have an actual propagating signal on record. Here we test this method on  models of collective behaviour in their deeply ordered phase by using ground truth data provided by numerical simulations in three dimensions. We compare two models  characterized by very different dynamical equations and information transfer mechanisms: the classic Vicsek model, describing an overdamped noninertial dynamics and the inertial spin model, characterized by an underdamped inertial dynamics. By using dynamic finite-size scaling, we show that spatio-temporal correlations are able to distinguish unambiguously the diffusive information transfer mechanism of the Vicsek model from the linear mechanism of the inertial spin model.

\end{abstract}

\maketitle

%\tableofcontents

%%%%%%%%%%%%%%%%%%%%%%%%%%
\section{Introduction}
%%%%%%%%%%%%%%%%%%%%%%%%%%

%{\bf Preamble}

Collective behaviour  is a widespread phenomenon in the living world, occurring over vastly different scales of space and time, and in a great variety of biological systems \cite{camazine_book,krause_book,sumpter_book}.  In recent years a strong interest has emerged in studying collective behaviour 
through the principles of statistical physics \cite{toner_review,ramaswamy_review,vicsek_review,marchetti_review}.
Following a paradigm typical of condensed matter, the first steps in this direction have been moved along two main paths: understanding how the motility of individuals, combined with the features of the interaction, determines the nature of the ordering transition  \cite{vicsek_review,chate+al_08} and studying what are the hydrodynamic properties at very large scales \cite{toner_review,marchetti_review}. 

A related, although distinct, question is that of how information propagates across the system and how it affects the collective response to perturbations. In many biological systems an efficient propagation of information across the group is a key to survival. Flocks of birds are a paradigmatic example in this respect: they are continuously subject to predatory attacks and yet they manage to respond very swiftly, changing collective direction of motion on very short timescales, still maintaining cohesion. This type of phenomenon suggests that the mechanism to transmit information across the group must be particularly efficient.

%experimental results

Recent experimental observations have shown that information propagates across flocks of starlings linearly and with very weak damping \cite{attanasi+al_14}. More precisely, the collective change of heading of a flock originates locally in space (from one bird) and it propagates to the rest of the flock as a wave, with a wavefront moving linearly in time with speed $c_s$. Linear information transfer is about the most efficient mechanism we can imagine. Former models models \cite{vicsek+al_95,gregoire+al_04} and hydrodynamic theories \cite{toner_review} fail to reproduce such behaviour, both at the analytic and at the numerical level \cite{cavagna+al_15}. 

In \cite{attanasi+al_14,cavagna+al_15} it has been formulated a novel theory which revolves around the concept of {\it behavioural inertia}, thus changing the dynamical differential equations for the velocities from first-order in time, to second-order. 
In its polarized phase -- the correct one to describe flocks -- the new theory reproduces  well the experimental data of bird flocks, showing that when inertia is taken into account the system can sustain information transfer through linear waves: if the heading of one individual changes, this change propagates quickly to the rest of the flock causing a collective turn, while keeping cohesion of the whole group; previous first-order models of flocks do not sustain linear information transfer, so that a change of heading of one individual led to global disruption of the group (see the simulations of \cite{cavagna+al_15}).
These results emphasize the relevance of second-order inertial terms in the dynamical equations of polarized systems, that is systems displaying long-range order. Even though inertial terms are asymptotically irrelevant over very large hydrodynamic scales \cite{toner+al_98,toner_review,ramaswamy_review}, they are in fact indispensable to explain the behaviour of finite-size real flocks \cite{cavagna2015silent}. 

%{\bf Current limitations}
It is therefore important to find a way, given an experimental data set, to assess whether inertial dynamics rules the system or not. 
For bird flocks this has been done through the direct experimental observation of the propagation of a wavefront across the group due to some spontaneous change of direction recorded on camera \cite{attanasi+al_14}. For large groups in the field this is, in general, a cumbersome way to proceed, as one needs to capture on record a collective change of state due to some uncontrolled perturbation. For example, in the study reported in \cite{attanasi+al_14} only 12 such events were captured in over four years of data-taking. Moreover, unlike in lab studies, in the field the possibilities to actively perturb the system are very limited. It seems therefore desirable to develop a more practical and effective approach to the problem.

%{\bf Objectives}

We discuss here a general method capable of learning whether or not inertial second-order dynamics rules a system directly from its unperturbed, spontaneous fluctuations. In this way we are able to tell whether or not a system is inertial by a sampling of its dynamics, rather than by manipulating it. The key tool of the method is  the spatio-temporal correlation function: the behavioural change of individual $i$ at time $t_0$ influences that of individual $j$ at a later time $t_0+t$. The form of this correlation, which extends in both space and time, bears the fingerprint of the dynamical equations ruling the system. We test this general method through numerical simulations of models with very different dynamical equations, namely with and without inertial terms. We find that spatio-temporal correlation successfully distinguish the different types of dynamics.

% Ordered phase

In the present work we will focus only on the {\it ordered} phase of the models of collective motion that we will analyze.
In the language of the renormalization group, this means that we will only be dealing with the 
zero temperature fixed point, rather than with the critical point. In biological terms we may say that the 
results we present are directly applicable only to polarized,`flock-like', groups, rather than unpolarized, `swarm-like', 
systems. Even though the analysis based on the spatio-temporal correlations has general validity, 
the results of such analysis (most importantly, the dynamical critical exponents) may depend on
such distinction. We will consider swarm-like systems in a future separate study.

%structure
The paper is organized as follows.
In Section II we introduce two archetypical model of self-propelled particles encoding non-inertial 
and inertial dynamics, and study their natural time scales on the basis of simple dimensional analysis. 
In Section III we define the spatio-temporal correlation function and discuss its general properties in Fourier space.
In Sec. IV we use dynamic finite-size scaling to predict how the correlation behaves in systems with finite size in the two cases of non-inertial and inertial dynamics. In this Section we also test our theoretical results against numerical simulations in three dimensions of the two models. 
In Sec.V we calculate the explicit form of the spatio-temporal correlation function under an approximate scheme. We discuss our conclusions in Section VI.

%%%%%%%%%%%%%%%%%%%%%%%%%%%%%%%%%%%%%%%%%
\section{Two different models of collective motion}
%%%%%%%%%%%%%%%%%%%%%%%%%%%%%%%%%%%%%%%%%

%%%%%%%%%%%%%%%%%%%%%%%%%%%%%%%%%%%%%%%%%
\subsection{Non-inertial dynamics: the Vicsek model}
%%%%%%%%%%%%%%%%%%%%%%%%%%%%%%%%%%%%%%%%%

The most important physics-inspired model of collective motion is the Vicsek model (VM) \cite{vicsek+al_95}. It describes a system of self-propelled particles with 
constant speed $v_0$, which interact through mutual imitation: each particle $i$ adjusts the direction of its velocity ${\bf v}_i$ by making it as close as possible to the mean direction of its neighbors. This effect of the neighbours is sometimes called {\it social force}. Several versions of the Vicsek model have been introduced and studied in the course of time (see, for example, \cite{gregoire2003moving, gregoire+al_04, chate+al_08, chate+al_08b}). 
Here we will consider its most basic form and, for the sake of analytic simplicity, we will write the dynamical equations in continuous time. 
In order to keep the speed $v_0$ of each particle constant, each velocity vector ${\bf v}_i$ is changed only by the component of the social force orthogonal to 
${\bf v}_i$. To this purpose, we introduce the following notation to indicate the projection of a generic vector $\bf w$ onto the plane orthogonal to ${\bf v}_i$,
\beq
{\bf w}^\perp \equiv {\bf w} - \left({\bf w}\cdot \frac{{\bf v}_i}{v_0}\right)\,\frac{{\bf v}_i}{v_0}  \ .
\eeq
We can thus write the Vicsek model in the following way,
\bea
\frac{\eta}{v_0} \frac{d {\bf v}_i}{dt} &=& \frac{1}{v_0}\left (J \sum_j n_{ij} {\bf v}_j \right)^{\perp}    +  {\boldsymbol{\zeta}}_i^{\perp}   
\label{vic1}
\\
\frac{d {\bf r}_i}{dt} &=&  {\bf v}_i \ .\label{vic2} 
\eea
where $\eta$ is a generalized friction coefficient and ${\boldsymbol{\zeta}}_i$ is a random white noise, whose variance is given by,
\begin{equation}
\langle {\boldsymbol{\zeta}}_i(t) \cdot {\boldsymbol{\zeta}}_j(t') \rangle= (2 d) \, \eta \, T \, \delta_{ij} \,  \delta(t-t') \ .
\label{vicnoise}
\end{equation}
We use the standard convention of expressing the amplitude of the noise as the product of the  `temperature' $T$ times the friction coefficient $\eta$ \cite{zwanzig_book}. 
Both the social force due to the neighbors and the noise contribute to changing the flight direction of the particle, but not its speed, thus ensuring that $|{\bf v}_i|=v_0$.
This formulation of the VM is slightly different from the one usually found in the literature, but fully equivalent. In particular, the friction $\eta$ could be eliminated in \eqref{vic1} and \eqref{vicnoise} 
by a rescaling of time \cite{zwanzig_book}, but this would make the comparison with the inertial model of the next Section (where $\eta$ cannot be rescaled away) much less transparent.

The adjacency matrix $n_{ij}$ is $1$ if $i$ and $j$ are interacting neighbors and $0$ if they are not, and it encapsulates the different kinds of interaction rules. If we consider, as in the original Vicsek model \cite{vicsek+al_95} {\it metric} interactions, then $n_{ij}=1$ if $r_{ij}\leq r_c$ and $n_{ij}=0$ if $r_{ij}> r_c$, where $r_c$ is the metric interaction range. If, on the other hand, interactions are {\it topological}, as it is the case in bird flocks \cite{ballerini+al_08}, then $n_{ij}=1$ if $j$ is within the first $n_c$ neighbors of $i$ and $n_{ij}=0$ otherwise, where $n_c$ is the topological interaction range. The adjacency matrix depends on time, $n_{ij}=n_{ij}(t)$: particles do not sit on a fixed lattice, they are self-propelled, so that the neighborhood of each particle evolves in time. This time-dependence is what makes self-propelled particles models intrinsically different from standard equilibrium statistical mechanics models.

Finally, we emphasize a very important fact: the Vicsek model has no inertial term. Even though one may be tempted to identify the l.h.s. of equation \eqref{vic1} as an inertial second-order term (it is, after all, an acceleration), this is not its correct interpretation. The fundamental degree of freedom of the model is the velocity, not the position, and indeed the social force at the r.h.s. of \eqref{vic1} is the derivative with respect to the velocity of a generalized Hamiltonian, function of the velocities, not of the positions, $H = - \sum_{ij} n_{ij} {\bf v}_i \cdot {\bf v}_j$. The overtones of ferromagnetic physics are evident. Hence, the Vicsek equation describes a non-inertial, first-order, overdamped dynamics for the velocity and for this reason the coefficient of $\dot {\bf v}_i$ is friction, not mass \cite{cavagna+al_15}.

%%%%%%%%%%%%%%%%%%%%%%%%%%%%%%%%%%%%%
\subsection{Inertial dynamics: the inertial spin model}
%%%%%%%%%%%%%%%%%%%%%%%%%%%%%%%%%%%%%

The Inertial Spin Model (ISM)  was introduced in \cite{attanasi+al_14} and in \cite{cavagna+al_15} to describe linear information transfer in natural flocks of birds. It correctly reproduces
 the way starling flocks perform collective turns and it provides the right dispersion law in these groups.
The ISM is described by the following equations,
\begin{subequations}
\begin{align}
\frac{d {\bf v}_i}{dt} &= \frac{1}{\chi} {\bf s}_i \times {\bf v}_i \label{ismvel}\\
\frac{d {\bf s}_i}{dt} &= \frac{{\bf v}_i}{v_0}
 \times \left[ \frac{J}{v_0} \sum_j n_{ij} {\bf v}_j - \frac{\eta}{v_0}\frac{d {\bf v}_i}{dt} + {\boldsymbol{\zeta}}_i \right] \label{ismspin}\\
\frac{d {\bf r}_i}{dt} &=  {\bf v}_i \ ,
\end{align}
\end{subequations}
with ${\bf v}_i \cdot {\bf s}_i = 0$.
The dynamic state of each particle is now described by two variables: the velocity ${\bf v}_i$ (again with fixed modulus $v_0$) and a new variable ${\bf s}_i$ called `spin'. As it can be seen from Eq.~(\ref{ismvel}), the spin describes how quickly the particle changes its direction of motion. The fact that, contrary to the Vicsek model, the social force $J\sum_j n_{ij}{\bf v}_j$ and the noise act on the spin, rather than directly on the velocity, indicates that the model is inertial, so that the instantaneous update of the velocities is smooth. The new parameter $\chi$ is the behavioral inertia; it is not the real mass, nor the mechanical moment of inertia. Rather, $\chi$ is an effective parameter describing the resistance of a bird to change the radius of curvature of its trajectory \cite{cavagna+al_15}.
If we take a further derivative of equation \eqref{ismvel} and exploit \eqref{ismspin}, we get a closed equation for the velocity,
\bea
\frac{\chi}{v_0} \frac{d^2 {\bf v}_i}{dt^2} + \chi \frac{{\bf v}_i}{v_0^3}\((\frac{d {\bf v}_i}{dt}\))^2 + \frac{\eta}{v_0} \frac{d {\bf v}_i}{dt} = 
\nonumber
\\
=\frac{1}{v_0} \left( J \sum_j n_{ij} {\bf v}_j\right)^{\perp} +{\boldsymbol{\zeta}}_i^{\perp}  \ .
\label{grand}
\eea
The ISM describes different dynamical regimes depending on the values of the parameters \cite{cavagna+al_15}. In particular, by tuning the friction $\eta$ with respect to the behavioural inertia $\chi$, we can explore both the overdamped regime (large $\eta^2/\chi$), where the model behaves as the Vicsek model, and the underdamped regime (small $\eta^2/\chi$), where the model displays the behaviour and dispersion law observed in natural flocks \cite{attanasi+al_14}. The Vicsek model of Eqs. (\ref{vic1}-\ref{vic2}) is exactly recovered when $\chi/\eta^2\to 0$. In the opposite limit, i.e. when $\eta^2/\chi \to 0$, we obtain instead a fully reversible dynamics, where the spins - which represent the generators of the rotational symmetry of the velocities - are strictly conserved quantities \cite{cavagna+al_15}.

%%%%%%%%%%%%%%%%%%%%%%%%%%%%%%%%%%%%%
\subsection{Natural time scales of the two models}
%%%%%%%%%%%%%%%%%%%%%%%%%%%%%%%%%%%%%

In order to work out the natural time scales of the dynamical equations introduced above,
it is convenient to introduce the (positive-definite) discrete Laplacian matrix,
\beq
\Lambda_{ij} = \delta_{ij}\sum_k n_{ik} \, - \, n_{ij} \ ,
\label{laplacian}
\eeq
which approximates in a discrete system the second-order derivative in space \cite{bollobas2013modern}. 
Notice that in a topological model the diagonal element of the
Laplacian is a constant, $\sum_k n_{ik} = n_c$, while in a metric model it fluctuates; in both cases, thus,  the amplitude of $\Lambda_{ij}$ 
is proportional to the mean number of interacting neighbours.
By using the discrete Laplacian we can rewrite the non-stochastic part
of the Vicsek models in the following way,
\beq 
\eta \frac{d {\bf v}_i}{dt} = \left (-J \sum_j \Lambda_{ij} {\bf v}_j   \right)^{\perp} \ ,
\eeq
where we have used the obvious relation: $({\bf v}_i)^\perp=0$. From this equation we can derive a time scale  by mere dimensional analysis. As we have said, the  Laplacian $\Lambda_{ij}$ is a discrete version of the second-order derivative in space, hence it is dimensionally equivalent to a term $k^2 a^2$, where $a$ is the
mean interparticle distance and $k$ is the momentum in Fourier space. We use a Fourier representation as this will be the space in which we will work for most of the paper. We have also seen that the discrete Laplacian is proportional to the mean number $n_c$ of interacting neighbours, so that, dimensionally, 
$\Lambda_{ij} \sim n_c a^2 k^2$. We therefore conclude that the natural time scale of the Vicsek model is given by,  
\beq
\tau^\mathrm{VM} \sim \frac{\eta}{J n_c a^2 k^2}  \ .
\label{zappo}
\eeq
This is a damping time scale expressing the collective relaxation time of the velocities in the Vicsek model. The fact that it
diverges for $k\to 0$ is a consequence of the fact that the theory has a diverging correlation length \cite{hohenberg1977theory}. 
The origin of this divergence and how it is tamed for finite size will be discussed in the next Section.

For the inertial spin model, equation \eqref{grand}, the situation is more complicated because this is a second-order equation and we expect to 
have two time scales, rather than one. As in the Vicsek case, the term $(\sum_i n_{ij} {\bf v}_i)^\perp$ can be substituted with the discrete Laplacian,
$(-\sum_i \Lambda_{ij} {\bf v}_i)^\perp$. Hence, the purely dimensional version of equation \eqref{grand} reads,
\beq
\chi/t^2+\eta/t -Jn_c a^2 k^2 = 0 \ .
\label{canasta}
\eeq
To simplify this expression it is convenient to introduce the reduced friction coefficient, $\gamma$, and the second sound speed,
$c_s$,
\beq
\gamma \equiv \frac{\eta}{2\chi} \quad , \quad c_s \equiv\sqrt{\frac{J n_c a^2}{\chi}}  \ .
\label{zappa}
\eeq
The parameter $c_s$ (which indeed has the physical dimensions of a velocity) is the speed of propagation of a signal in the inertial case \cite{cavagna+al_15}.  Using $\gamma$ and $c_s$ into \eqref{canasta} and multiplying by $t^2$, we get,
\beq
(c_s k)^2 \; t^2 - \gamma \, t- 1=0  \ ,
\label{pongo}
\eeq
which, on merely dimensional grounds, provides the two obvious time scales,
\beq
 \tau_1^\mathrm{ISM} \sim \frac{1}{\gamma}  \quad , \quad \tau_2^\mathrm{ISM} \sim \frac{1}{c_sk}  \ .
\label{zappe}
\eeq
The damping scale is the one containing the friction coefficient, that is $\tau_1^\mathrm{ISM}$: larger friction $\gamma$ (but not `too large', otherwise one recovers the Vicsek model -- see below) produces a shorter (quicker) 
damping time. As expected, this dissipative time scale accompanies the linear term in $t$, which breaks the time-reversal symmetry.
At variance with the non-inertial case, the damping time is now independent of $k$. 
The second time scale of the inertial case contains the signal propagation speed, $c_s$, hence it is naturally associated to the period 
of oscillation of mode $k$. Indeed, when the friction is small, $\gamma \ll c_s k$, that is when $\tau_1^\mathrm{ISM} \gg \tau_2^\mathrm{ISM}$,  
the ISM equation becomes time-reversible and propagation emerges \cite{cavagna+al_15}.
It is important to notice that the inertial time scale $\tau_2^\mathrm{ISM}$ does depend on $k$, although with  exponent $1$, rather than $2$ as in the non-inertial case. This linear dependence on $k$ is responsible for linear information transfer in the inertial case \cite{cavagna+al_15,cavagna2015silent}.

In the limit of large damping and low inertia we expect to recover the Vicsek model. This overdamped limit is obtained when 
the signal gets damped before it can propagate, which corresponds to having $\gamma \gg c_s k$, that is $\tau_1^\mathrm{ISM} \ll \tau_2^\mathrm{ISM}$. Notice that this limit can be also written as, $k \ll \gamma/c_s$; the overdamped limit is therefore a limit of small momentum $k$ and it is in fact nothing else than the {\it hydrodynamic limit}, in which inertia is always sub-dominant with respect to damping \cite{zwanzig_book}. If we rewrite \eqref{pongo} as,
\beq
\left(t/\tau_2^\mathrm{ISM} \right)^2 - t/\tau_1^\mathrm{ISM} - 1=0 \ ,
\label{pongus}
\eeq
we see that when $\tau_1^\mathrm{ISM} \ll \tau_2^\mathrm{ISM}$, the discriminant of this equation simplifies and we are left with just one time scale, namely $(\tau_2^\mathrm{ISM})^2/\tau_1^\mathrm{ISM} = \eta/(Jn_c a^2 k^2)$, which is the same time scale of the Vicsek model, as expected.

The important conclusion of this Section is that the two different dynamics are ruled {\it at the naive dimensional level} by time scales that have a very different dependence on $k$. The time scale of the non-inertial Vicsek case decays as $1/k^2$, while in the inertial case the period goes as $1/k$ and the damping time does not depend on $k$. These exponents  need not to be exact, of course, as this was mere dimensional analysis, and they can change because of renormalization and off-equilibrium effects \cite{toner_review}. However, dimensional analysis is helpful, as it unveils a deep, intrinsic difference between the natural time scales of the two models, suggesting that it should be quite possible to detect such difference in the data, both at the numerical and at the experimental level. We will see in the next Sections how to practically do that.

%%%%%%%%%%%%%%%%%%%%%%%%%%%%%%%%%%%%%%%%%%%%%
\section{The spatio-temporal correlation function}
%%%%%%%%%%%%%%%%%%%%%%%%%%%%%%%%%%%%%%%%%%%%%

We will introduce in this Section the velocity correlation function in space and time, which will be our main tool of analysis. 
Before entering into the details, though, we need to discuss a preliminary issue.

\subsection{The issue of anisotropy}

As we stated in the Introduction, we will deal in this work with systems in their ordered phase, that is systems in which the rotational symmetry is spontaneously broken. In this case two kind of anisotropies arise in the correlation, and we shall briefly discuss them here. 
% v_perp vs v_parallel
First, the velocity field has far stronger fluctuations in the transverse than in the longitudinal direction. 
\footnote{In what follows, we call {\it longitudinal} the mean direction of motion of the flock, whereas 
{\it transverse} are all directions belonging to the plane orthogonal to the mean direction of motion.}
This is a well-known fact, also present in equilibrium systems \cite{patashinskii_book}; it implies that when one computes the correlation of the full velocity vector ${\bf v}_i$, one is actually measuring the transverse correlation. In fact, transverse and longitudinal fluctuations are connected to each other and it was first demonstrated in \cite{pata1973chi_longi} that the longitudinal susceptibility in the broken-symmetry is slave to the transverse susceptibility, $\chi_\parallel \sim \chi_\perp^{1/2}$. Non-equilibrium effects induced by the self-propulsion change this exponent, as recently discovered in \cite{ginelli2016chi_longi}, but the fact remains that longitudinal fluctuations are highly suppressed with respect to transverse ones. 
% k_perp vs k_parallel
Secondly, it has been demonstrated in \cite{toner+al_98, toner1998flocks} that in active systems with spontaneously broken symmetry, the feedback between velocity and position produces also a spatial anisotropy in the decay of the correlation; more precisely, the critical exponents along the longitudinal and transverse directions are different from each other. This effect is absent in equilibrium systems and it is therefore quintessential of active matter; in two dimensions it explains the suppression of fluctuations and the stabilization of long-range order \cite{toner+al_98, toner1998flocks}.

It therefore seems that to give a thorough characterization of the correlation in an active system one should use the anisotropic pairs, $({\bf v}_\parallel,{\bf v}_\perp)$ and $({\bf k}_\parallel, {\bf k}_\perp)$, rather than the full isotropic vectors, $\bf v$ and $\bf k$. A detailed anisotropic characterization, though, is not our aim here. Rather, what we want to do is to distinguish different types of dynamics (inertial vs non-inertial), from the correspondingly different forms of the correlation function. As we shall see, an isotropic study, which does not distinguish between longitudinal and transverse directions, is sufficient to this purpose, as it provides a very strong signal able to pick up different dynamical rules. This is fortunate, as in most real biological data sets it is quite difficult to have enough statistics to be able to separate the correlation into longitudinal and transverse components \cite{cavagna+al_10}. Considering that we want to set a backdrop for future comparisons between theory and simulations on one side, and experiments on the other, we believe that an isotropic study is a first necessary step to take in order to distinguish inertial from noninertial systems.

%%%%%%%%%%%%%%%%%%%%%%%%%%%%%%%%%%%%%%%%%%
\subsection{The connected correlation function}
%%%%%%%%%%%%%%%%%%%%%%%%%%%%%%%%%%%%%%%%%%

The connected velocity correlation, $C_{ij}$, between individuals  $i$ and $j$, measures how much a change of direction
of $i$ influences (and is influenced by) a change of direction of $j$, 
\beq
C_{ij}\equiv\langle \delta {\bf v}_i \cdot \delta {\bf v}_j\rangle \ ,
\label{minkia}
\eeq
where the velocity fluctuation $\delta {\bf v}_i$ is  the deviation of the velocity of $i$ from the mean velocity of the group, 
\beq
\delta {\bf v}_i \equiv {\bf v}_i - \frac{1}{N}\sum_k {\bf v}_k \ .
\eeq
The systems we are studying are in the polarized phase, hence the mean velocity is strongly different from zero; it would therefore make 
no sense to compute the correlation of the full velocities (non-connected correlation), as this would be completely dominated by the mean velocity; the only physically and biologically relevant correlation is that between the fluctuations.
However this fluctuation is a dimensional quantity; therefore, in order to have a more suitable quantity to make a comparison between natural and numerical systems, we prefer to work with the dimensionless velocity fluctuation,
\beq
\delta \hat{\bf v}_i \equiv \frac{\delta{\bf v}_i}{\sqrt{\frac{1}{N} \sum_k \delta{\bf v}_k \cdot \delta{\bf v}_k}}  \ .
\label{fluctu}
\eeq
We now need to define the connected correlation in both time and space, between
individual $i$ at time $t_0$ and individual $j$ at time $t_0+t$.
Following van Hove \cite{van1954correlations, hansen_book}, we 
define the  spatio-temporal correlation function as, 
\begin{widetext}
\beq
C({\bf r},t) = \left\langle  \frac{1}{N \rho} \sum_{i,j}^N  \delta\hat{\bf v}_i(t_0) \cdot \delta\hat{\bf v}_j(t_0+t) \, 
\delta^{(3)}[{\bf r} - {\bf r}_i(t_0)+{\bf r}_j(t_0+t)] \right\rangle_{t_0} .
\label{mingus}
\eeq
\end{widetext}
The positions are calculated with respect to the center of mass of the system, that is ${\bf r}_i(t_0) = {\bf R}_i(t_0) -{\bf R}_{\mathrm{CM}}(t_0)$ (and similarly 
for $j$); 
$\rho$ is the density; the bracket indicates an average over time,
\beq
\langle f(t_0, t)\rangle_{t_0}=\frac{1}{t_\mathrm{tot}-t} \sum_{t_0=1}^{t_\mathrm{tot}-t} f(t_0, t) \ ,
\eeq
and $t_\mathrm{tot}$ is the total available time in the simulation or in the experiment. 
It is important to remark that in \eqref{mingus} everything related to  particle $i$ (position and velocity) is evaluated at time $t_0$, while
everything related to $j$ is evaluated at time $t_0+t$. The basic idea of \eqref{mingus} is to sum the correlation of all pairs that 
over time $t$ are found at distance $\bf r$ from each other.
Note that, thanks to the normalization $1/\rho$, the spatio-temporal
correlation function defined in \eqref{mingus} is dimensionless.
We can define the scalar distance between $i$ and $j$ at different times, 
\beq
r_{ij}(t_0,t)\equiv |{\bf r}_i(t_0)-{\bf r}_j(t_0+t)|
\eeq
and rewrite the correlation function as,
\begin{widetext}
\beq
C(r,t) = 
\left\langle  \frac{1}{N 4 \pi r^2 \rho}
\sum_{i,j}^N  \delta\hat{\bf v}_i(t_0) \cdot \delta\hat{\bf v}_j(t_0+t) \, 
\delta[r - r_{ij}(t_0,t)] \right\rangle_{t_0}  \ .
%=
%\left\langle  \frac{\sum_{i,j=1}^N  {\bf v}_i(t_0) \cdot {\bf v}_j(t_0+t) \, 
%\delta(r - r_{ij}(t_0,t))}{\sum_{i,j=1}^N \delta( r - r_{ij}(t_0,t))}\right\rangle_{t_0}
\eeq
\end{widetext}
Notice that for $t=0$ this quantity reduces to the standard static correlation function previously used in the studies of bird flocks \cite{cavagna+al_10} and insect swarms \cite{attanasi2014finite, attanasi2014collective}, because the normalization term, $N 4 \pi r^2 \rho$,  represents the number of pairs at distance $r$, i.e. $\sum_{i,j} \delta (r - r_{ij})$. The purpose of $C(r,t)$ is to measure how much a  change of velocity of an individual at time $t_0$ influences a change of velocity of another individual at distance $r$ at a later time $t_0+t$.

%%%%%%%%%%%%%%%%%%%%%%%%%%%%%%%%%%%%%
\subsection{The correlation function in Fourier space}
%%%%%%%%%%%%%%%%%%%%%%%%%%%%%%%%%%%%%

In order to calculate the collective time scale of the system, namely the longest time scale of relaxation, one normally
introduces the space-integral of $C(r,t)$, 
\beq
C(t) = \rho \int d{\bf r}\ C(r,t)  \ ,
\label{palito}
\eeq
so as to remain with a purely time-dependent correlation function (we multiply the r.h.s. by a factor $\rho$ in order to have the same physical dimensions for $C(t)$ and $C(r,t)$). However, this prescription cannot work for us, and in general for discrete data in active systems.
Remember our definition of velocity fluctuations, equation \eqref{fluctu}: we subtract to each velocity the {\it space} average, rather than the {\it ensemble}
average. In fact, we have no choice: in active systems individuals move, do not sit on a lattice, hence we need to define averages at any fixed time, and this leaves us with just one choice, the space average. As a result, we inherit the sum rule,
\beq
\sum_i \delta \hat{\bf v}_i =0 \quad - \quad \mathrm{sum \ rule}  \ ,
\label{sumrule}
\eeq
whose main consequence is the rather unfortunate result, 
\beq
C(t) =  \frac{1}{N} \sum_{i,j}^N  \delta\hat{\bf v}_i(t_0) \cdot \delta\hat{\bf v}_j(t_0+t) = 0 \ .
\label{disapp}
\eeq
Hence, the sum rule \eqref{sumrule} prevents us from defining a collective time correlation integrated in space.
This problem can be circumvented by defining the correlation function in Fourier space,
\begin{widetext}
\beq
C(k, t) = \rho \int d{\bf r} \ e^{i {\bf k}\cdot{\bf r}} C(r,t)
=
\rho \int dr \; 2\pi r^2 \, C(r,t) \int_{-1}^{1} d(\cos\theta)\ e^{i k r \cos \theta} \ .
\eeq
\end{widetext}
The integral over $\cos\theta$ can be performed explicitly and the Dirac's delta used to collapse the integral in $dr$, so that we finally obtain,
%\begin{widetext}
\beq
C(k, t) = 
\left\langle \frac{1}{N} \sum_{i,j}^N   \frac{\sin(k r_{ij}(t_0,t))}{k \, r_{ij}(t_0,t)} \, \delta\hat{\bf v}_i(t_0) \cdot \delta\hat{\bf v}_j(t_0+t)  \right\rangle_{t_0}  .
\label{dangle}
\eeq
%\end{widetext}
This quantity is far more useful than the mere space integral \eqref{palito}, as by varying the momentum $k$ we can integrate the correlation over different length scales, making it a useful tool even in presence of the sum rule \eqref{sumrule}. Moreover, it can be easily measured from the data, both in numerical simulations and in experiments. In the rest of this work 
we plan to use the spatio-temporal correlation $C(k,t)$ defined in \eqref{dangle} as the principal instrument of analysis to distinguish inertial from non-inertial dynamics.

%%%%%%%%%%%%%%%%%%%%%%%%%%%%%%%%%%%%%
\subsection{The static limit of the correlation function}
%%%%%%%%%%%%%%%%%%%%%%%%%%%%%%%%%%%%%

A particularly important case is the static limit of the correlation function \eqref{dangle}, namely the case $t=0$, in which we recover the 
Fourier transform of the static correlation function, 
\beq
C_0(k)\equiv C(k,t=0) = 
\left\langle \frac{1}{N } \sum_{i,j}^N \;  \frac{\sin(k\,r_{ij})}{k\,r_{ij}} \ \delta\hat{\bf v}_i \cdot \delta\hat{\bf v}_j \, \right\rangle_{t_0} \ ,
\label{galore}
\eeq
where now both $i$ and $j$ are evaluated at equal time $t_0$. What can we say, on general grounds, about the behaviour of $C_0(k)$? 
For $k\to0$ we have $\sin (kr_{ij})/kr_{ij}\to1$ and the correlation therefore 
is zero at all times, due to the sum rule \eqref{sumrule},
\beq
\lim_{k\to 0}C_0(k)=0  \quad - \quad \mathrm{sum \ rule} \ .
\label{ugurii}
\eeq
On the other hand, for a generic nonzero value of $k$, the sum in \eqref{galore} is dominated by particle pairs with distance
$r_{ij} < 1/k$, as larger distances are suppressed by the factor $\sin (kr_{ij})/kr_{ij}$ becoming a rapidly oscillating term.
In particular, for $k \to \infty$ all terms with $r_{ij}\neq 0$, and therefore with
$i \neq j$, get killed in the sum and only the terms $i=j$ survive. This implies that,
\beq
\lim_{k\to \infty} C_0(k) = 
\left\langle \frac{1}{N} \sum_{i}^N \;  \delta\hat{\bf v}_i ^2 \right\rangle_{t_0} =1 \ ,
\label{self}
\eeq
because of definition \eqref{fluctu} of the dimensionless velocity fluctuations. 

To make further progress we need to introduce a new crucial player, namely 
the correlation length, $\xi$, which is a measure of the size of the correlated regions in the system \cite{binney_book}.
When the momentum $k$ is decreased from $k=\infty$, but it is still larger than $1/\xi$, the factor $\sin (kr_{ij})/kr_{ij}$ is
dominated by pairs with $r_{ij} <\xi$: hence, by decreasing $k$ we are adding in the sum \eqref{galore} more and more
correlated pairs. We therefore expect $C_0(k)$ to {\it increase} when $k$ is decreased from $k=\infty$. When the
momentum arrives at $k\sim 1/\xi$, we have added to the sum in \eqref{galore} all correlated pairs (that is pairs within one correlation 
length) and we start adding uncorrelated pairs, which lie beyond $\xi$. Hence, we do not expect  $C_0(k)$ to further increase on
decreasing $k$ below $1/\xi$. In fact, if we performed ensemble averages, $C_0(k)$ would level; however, we perform spatial
averages and we are bound by the sum rule \eqref{ugurii}.
As a consequence, decreasing $k$ below $1/L$ (where $L$ is the system's size) has the effect to {\it decrease} the static correlation $C_0(k)$, until eventually it vanishes for $k=0$. We therefore expect the static correlation to have a maximum. In a generic system, where there is no relation between $\xi$ and $L$, the position of this maximum is a complicated function of these two scales (e.g. if $\xi \ll L$ the static correlation has -- in log scale -- a broad plateau between $k\sim1/\xi$ and $k\sim1/L$). However, polarized flocks, which are the object of the present study, are scale-free systems, where $\xi\sim L$ (see next section); in this case, the maximum of $C_0(k)$ occurs unambiguously at $k_\mathrm{max} \sim 1/\xi \sim 1/L$. This behaviour of the static correlation $C_0(k)$ is confirmed by numerical simulations, see figures 1(a) and 2(a).

In statistical physics, the static correlation at $k=0$, that is the volume integral of the correlation in $\bf r$ space, is the 
susceptibility, $\chi_\mathrm{stat}$, of the system \cite{binney_book} (we use the `stat' subscript to distinguish this quantity from the 
generalized inertia of the ISM equations). In our case, though, we have $C_0(k=0)=0$ by construction, due to sum rule
\eqref{sumrule}, so this cannot be the susceptibility.
However, we have  seen that $C_0(k)$ has a maximum at intermediate momentum, 
$k=k_\mathrm{max}\sim 1/\xi$. The value of the static correlation at this maximum, $C_0(k_\mathrm{max})$, is equal to the integral of the static 
real-space correlation up to $r\sim \xi$. Hence, the fact that $C_0(k)$ peaks at $k_\mathrm{max}$ 
indicates that $C_0(k_\mathrm{max})$ is a fair estimate of the static susceptibility, 
\beq
\chi_\mathrm{stat} \sim C_0(k_\mathrm{max}) \quad , \quad \xi \sim 1/k_\mathrm{max}  \ .
\eeq
Notice that evaluating the susceptibility in this way is equivalent to performing the space integral 
of the real space correlation $C_0(r)$ up to the point where this integral peaks, which is what has been
done in \cite{attanasi2014finite}. 

To conclude this Section we notice that in the static limit the Vicsek model and the inertial spin model {\it must} have the same behaviour, and therefore the same static correlation function $C_0(k)$. This is a very general consequence of the fact that friction always wins over 
inertia at steady state \cite{zwanzig_book}. We will show in numerical simulations that this is indeed the case.

%%%%%%%%%%%%%%%%%%%%%%%
\section{Finite size scaling}
%%%%%%%%%%%%%%%%%%%%%%%

Both numerical and real biological systems have finite size. In this Section we will discuss how to take care of finite-size effects and how to exploit them to our advantage in distinguishing inertial from non-inertial dynamics. From now on we will only deal with collective systems in their ordered (i.e. polarized) phase. In physical terms this means that we are far from the ordering transition (be it in temperature, or in density). In biological terms, this means dealing with flocks, rather than swarms. 
For a finite-size scaling analysis of the dynamics of the Vicsek model close to the 
ordering transition (low polarization) we refer the reader to the comprehensive work of Ba\-glietto and Albano 
\cite{baglietto2008finite}.

%%%%%%%%%%%%%%%%%%%%%%%%%%%%%%%%%%%%%
%%%%%%%%%%%%%%
\subsection{Static correlation}
%%%%%%%%%%%%%%
%%%%%%%%%%%%%%%%%%%%%%%%%%%%%%%%%%%%%

In order to make some predictions about the form of the static correlation function, $C_0(k)$, we start from
the classic scaling hypothesis of Widom and Kadanoff \cite{widom1965equation,kadanoff1966introduction},
which states that, for length scales much larger than the interparticle distance $a$, namely for $k \ll 1/a$, and for large
correlation length, $\xi \gg a$,
the correlation function depends on the control parameters of the theory only through the correlation length:
\beq
C_0(k) = 
\frac{1}{k^{\gamma/\nu}} \hat f_0(k \xi) \ ,
\label{widom}
\eeq
where $\gamma/\nu$ is the ratio between the susceptibility and correlation length critical exponents  \cite{binney_book};
$\hat f_0$ is a scaling function whose explicit form will depend on several nonuniversal factors, including the particular choice of how we
perform the statistical averages (space average vs ensemble average). The homogeneous form \eqref{widom} of
the correlation function is one of the cornerstones of the theory of critical phenomena: once we observe the system over scales
much larger than the discrete microscopic spatial mesh (that is scales larger than the lattice spacing, or interparticle distance, $a$), the only relevant length 
scale that remains in the theory is the correlation length, $\xi$. The general aim of a finite-size scaling analysis is to learn the value of the
critical exponent $\gamma/\nu$, without the need to know the explicit form of the nonuniversal scaling function $\hat f_0$.

%%%%%%%%%%%%%%%%%%%%%%%%%%%%%%%%%%%%%%%%%%%%%%%%%%%
\clearpage

\onecolumngrid

\begin{figure}[!h]
\centering
\includegraphics[width=0.7\textwidth]{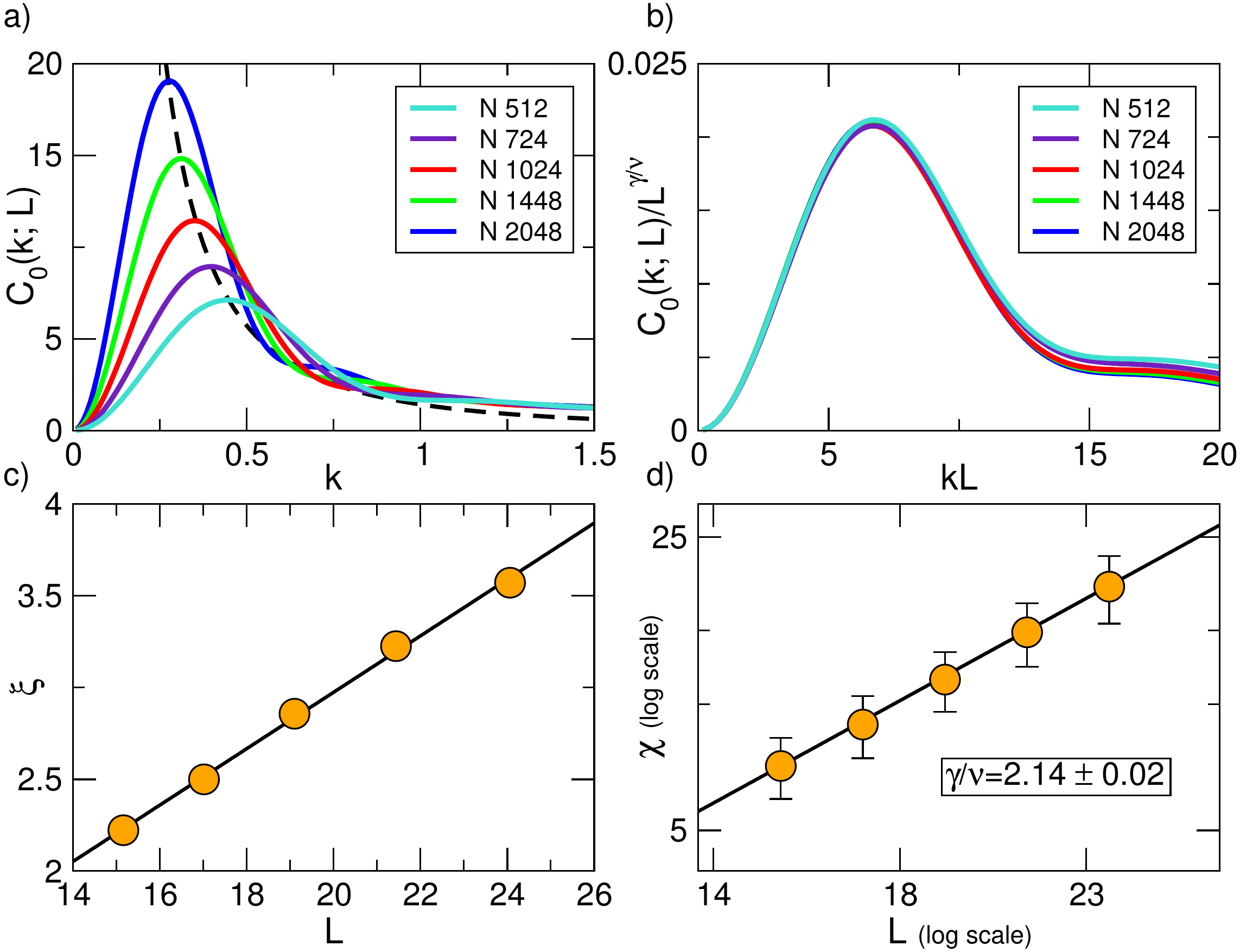}
\caption{{\bf Static behaviour of the non-inertial model.} Numerical simulation of the Vicsek model in $d=3$ with topological interaction (metric interaction gives identical results).
a) Static correlation as a function of the momentum at various values of the size $L$, displaying a clear maximum at $k=k_\mathrm{max}$. The position and
the height of the maximum shift with $L$, giving rise to a finite-size proxy of the bulk divergence (dashed line). b) Rescaled correlation function, according to
equation \eqref{muzio}. The best collapse occur for $\gamma/\nu= 2.14$. c) Correlation length, defined as $\xi \sim 1/k_\mathrm{max}$ as  function of the size $L$ (linear scale). d) Susceptibility, defined as $C_0(k_\mathrm{max}(L);L)$, as a function of the size (log-log scale).
}
\end{figure}

\twocolumngrid

\noindent 
%%%%%%%%%%%%%%%%%%%%%%%%%%%%%%%%%%%%%%%%%%%%%%%

\noindent
There is a second ingredient we need to take into account to make progress.
Flocks, and polarized natural groups in general, are systems with spontaneously broken continuous symmetry (rotation). For this reason
we expect to find some off-equilibrium relic of the Goldstone theorem \cite{goldstone1961field}, which states that the static 
correlation function must be scale-free (or long-range)  \cite{patashinskii_book, ryder1996quantum}. 
In systems with finite size $L$, this scale-free condition implies that the only intrinsic length scale of the system is the size of 
the system itself, namely,
\beq
\xi \sim L  \ .
\label{nemo}
\eeq
Relation \eqref{nemo} has been consistently verified at the numerical \cite{chate+al_08b} and theoretical \cite{toner_review} 
level for the Vicsek model, and at the experimental level for bird flocks \cite{cavagna+al_10}.
In the scale-free case the Widom-Kadanoff scaling form \eqref{widom} becomes, 
\beq
C_0(k;L) = 
\frac{1}{k^{\gamma/\nu}} \hat f_0(k L) \ .
\label{muzia}
\eeq
The scaling function obeys the relation \cite{widom1965equation},
\beq
\lim_{x\to\infty}\hat f_0(x) = \mathrm{constant} \ ,
\eeq
and therefore, in the infinite-size limit, the correlation function takes the simple power-law form, 
\footnote{
We emphasize again that all these scaling relations only hold for wavelengths long compared to the interparticle distance, 
$k \ll 1/a$. For $k > 1/a$, on the other hand, the scaling form \eqref{muzia} is not valid, so that we cannot perform the 
limit $k\to\infty$ in it; for this reason the limit for $k\to \infty$ of $C_0$ is $1$, as expressed by equation \eqref{self},
rather than $0$, as \eqref{muzia} would seem to suggest.}
\beq
C_0(k) \sim \frac{1}{k^{\gamma/\nu}} \quad \ , \quad L=\infty \ .
\label{diverge}
\eeq
The divergence for $k\to 0$ implies that not only the correlation length, but also
the static susceptibility, $\chi= C_0(k=0)$, is infinite in the $L=\infty$ limit of a scale-free system \cite{binney_book}. 
Relation \eqref{diverge} holds irrespective of how we perform the statistical averages, 
because in a very large system the correlation function is a self-averaging quantity;
this is why $C_0(k)$ does not depend on the scaling function $\hat f_0$ in the bulk. 
A naive dimensional analysis of both the Vicsek and the ISM equation gives,
\beq
(\gamma/\nu)_\mathrm{naive}=2  \ ,
\eeq 
because the discrete Laplacian simply translate into a $k^2$ term. However, naive dimensional analysis could be corrected by
off-equilibrium effects \cite{toner_review}, therefore we hold no prejudices on the true value of the exponent $\gamma/\nu$.

\clearpage

\onecolumngrid

%%%%%%%%%%%%%%%%%%%%%%%%%%%%%%%%%%%%%%%%%%%%%%%%%%%
\begin{figure}[!h]
\centering
\includegraphics[width=0.7\textwidth]{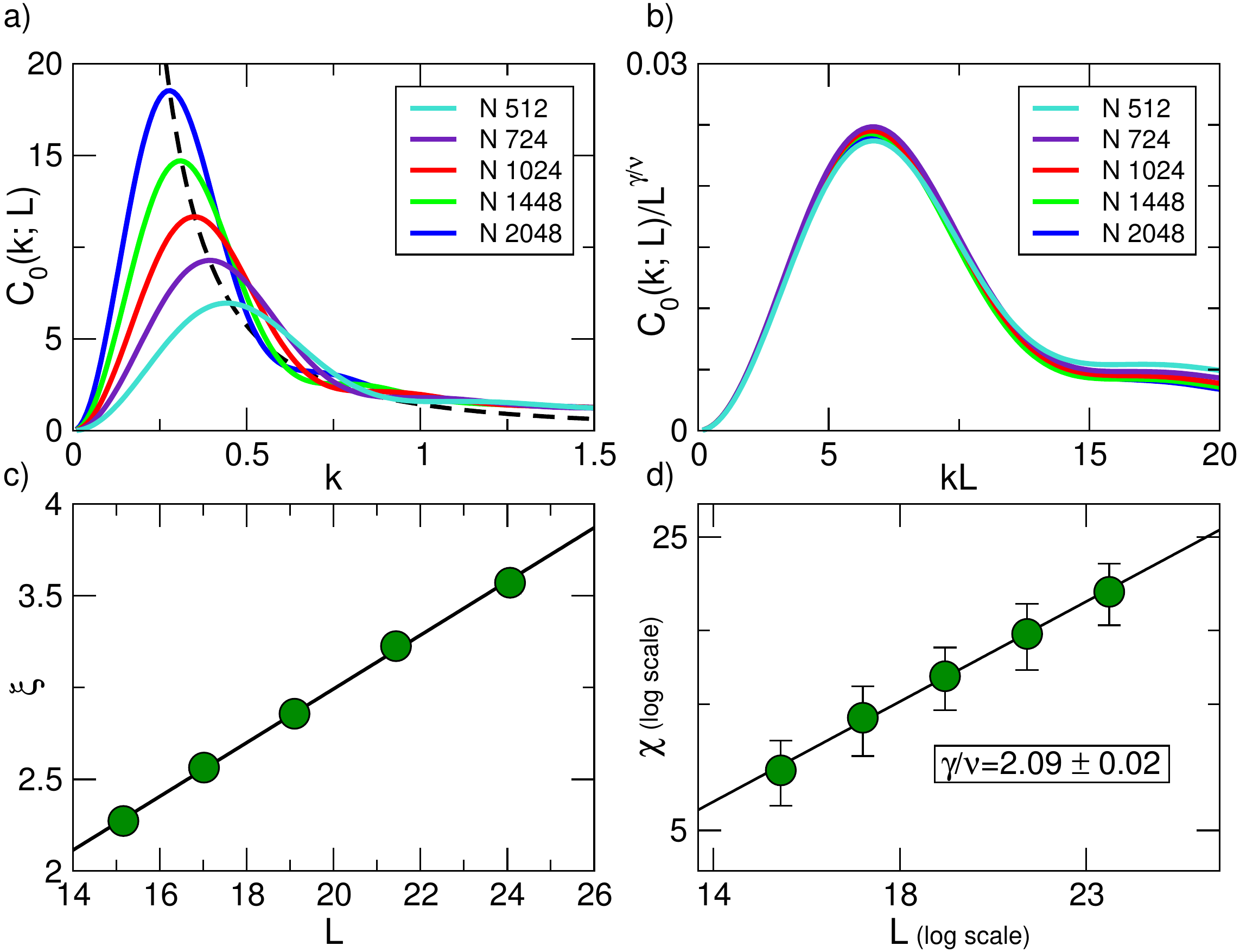}
\caption{{\bf Static behaviour of the inertial model.} Numerical simulation of the inertial spin model in $d=3$ with topological interaction (metric interaction gives identical results).
All panels are the same as in Fig. 1. The best collapse occur for $\gamma/\nu=2.09$.
}
\end{figure}
%%%%%%%%%%%%%%%%%%%%%%%%%%%%%%%%%%%%%%%%%%%%%%%
\twocolumngrid

\noindent 
Going back to finite-size systems, we can multiply and divide equation \eqref{muzia} by $L^{\gamma/\nu}$, to obtain the following equivalent scaling 
form of the correlation function,
\beq
C_0(k;L)  =  L^{\gamma/\nu} f_0(kL)  
\ ,
\label{muzio}
\eeq
which is the one we shall test in numerical simulations.
From the discussion of  the previous Section, we expect $C_0(k;L)$ to have a maximum as a function of $k$ at,
\beq
k_\mathrm{max}\sim 1/\xi \sim 1/L \ .
\label{ciupa}
\eeq
Moreover, as we have seen the finite-size susceptibility is given by the value of the static correlation at its maximum, 
\beq
C_0(k_\mathrm{max};L) \sim \chi_\mathrm{stat} \sim L^{\gamma/\nu} \ ,
 \label{nema}
\eeq
which is the standard finite-size behaviour of the susceptibility in critical phenomena.
\footnote{Equation \eqref{nema} is only valid when the density and interaction range are kept constant on changing $L$, as in the current work. When these quantities vary one must keep into account some corrections, which are particularly relevant in the metric interaction case; this point is carefully explained in the SI of \cite{attanasi2014finite}.}

We have checked all the predictions of this Section by performing numerical simulations in $d=3$
in the ordered phase (polarization $\Phi=0.9$) of both the Vicsek model and of the inertial spin model 
(details of the numerics are provided in Appendix A). We have computed in the two cases the static correlation function
in Fourier space, $C_0(k,L)$, according to its definition, equation \eqref{dangle}, at various values of the size $L$. 
Results are reported in Figs.1 and 2.

First, we observe that the numerical results are virtually indistinguishable between
Vicsek model and ISM: the static phenomenology of the two models is the same, as it should be. This is a nontrivial
consistency check; in particular, because the dynamical time scales of the two models are very different, obtaining the
same static behaviour is an indication that the simulation has thermalized for both models, so that we are really observing
steady-state results, rather than transients. Of course, this result also clearly shows that in order to distinguish the two models
we need to go beyond the purely static correlation, which is what we shall do in the following Sections.

To test the static scaling form \eqref{muzio} we notice that this relation implies 
that by plotting $C_0(k; L)/L^{\gamma/\nu}$ vs $kL$, correlations
calculated at different sizes $L$ should all collapse on the same curve, provided that we
find the correct value of the exponent $\gamma/\nu$. This is exactly what 
happens (panel (b) of Figs.1 and 2).
The collapse of the correlation functions at different sizes provides an estimate of the exponent $\gamma/\nu$. 
The data show that this static scaling exponent is very close to $2$, the value predicted by  naive
dimensional analysis; given that the two models have identical static behaviour, we averaged their exponents, thus obtaining, 
\beq
(\gamma/\nu)_{d=3} = 2.11 \pm 0.02  \ .
\eeq
Hence, it seems that, at least in this case of highly ordered phase ($\Phi=0.9$), off-equilibrium corrections are weak.
We notice that this value of the polarization $\Phi$ is not at all uncommon in real biological systems, as bird flocks \cite{cavagna+al_10}, 
\cite{lukeman2010inferring}, and fish schools \cite{gautrais2012deciphering}. Notice also that we find a consistent scaling phenomenology up to $N=2048$
particles, a number definitely not too small for many biological groups. Hence, naive dimensional analysis seems rather 
robust in the ordered phase of such medium-large systems. We will see that this is also the case at the dynamical level.

Relations \eqref{ciupa} and \eqref{nema} regarding the scaling of the correlation length and of the susceptibility are equally well
satisfied by the numerical data (panels (c) and (d) of  Figs.1 and 2). Moreover, if we evaluate the static correlation at $k_\mathrm{max}$ and plot it as a function 
of $k_\mathrm{max}$ (rather than $L$), we obtain from \eqref{muzio} the pseudo-bulk relation,
\beq
C_0(k_\mathrm{max}(L);L)  \sim \frac{1}{k_\mathrm{max}^{\gamma/\nu}} \ ,
\label{nusco}
\eeq
which is the finite-size representation of the infinite-size behaviour \eqref{diverge}. Equation \eqref{nusco} is reported as a dashed line 
in Figs.1a and 2a. 

The results of this Section show that the classic static scaling relations of Widom and Kadanoff, and the scale-free 
consequences of Goldstone's theorem, are verified quite accurately in both models of collective motion.

%%%%%%%%%%%%%%%%%%%%%%%%%%%%
\subsection{The dynamic scaling hypothesis}
%%%%%%%%%%%%%%%%%%%%%%%%%%%%

In order to study the fully dynamical correlation function we turn to a powerful concept in classical statistical mechanics, namely
that of {\it dynamic scaling}.
The dynamic scaling hypothesis was formulated by Halperin and Hohenberg in \cite{HH1967scaling} and \cite{HH1969scaling}
as a generalization of the static scaling relation of Widom and Kadanoff, equation \eqref{widom}. The dynamic 
scaling hypothesis makes two assertions: i) the characteristic time scale of the spatio-temporal correlation (or, equivalently, its characteristic frequency)
 is a homogeneous function of the momentum $k$ and of the correlation length $\xi$,
\beq
\tau= \frac{1}{k^z} h(k\xi) \ ,
\label{hh1}
\eeq
a relation that defines the so-called {\it dynamical critical exponent}, $z$; ii) the dynamical part of the spatio-temporal correlation is a function of the product $k\xi$, rather than 
of these two variables independently, 
\beq
C(k,t; \xi) =C_0(k;\xi)\;  f\left(\frac{k^zt}{h(k\xi)}, k\xi \right) \ .
\label{hh2}
\eeq
The dynamic scaling hypothesis is a concept deeply rooted in the renormalization group idea and it basically states that the correlation length is 
the only relevant length scale in a system, even at the dynamical level, so that the product $k\xi$ is the only way external tuning parameters (as temperature and 
noise) may enter the spatio-temporal correlation \cite{HH1969scaling}. In the rest of this Section we will use the dynamic scaling hypothesis to distinguish inertial from noninertial model.

%%%%%%%%%%%%%%%%%%%%%%%%%%%%
\subsection{Dynamical correlation: Non-inertial case}
%%%%%%%%%%%%%%%%%%%%%%%%%%%%

In Section II we have derived through naive dimensional analysis the time scale of the Vicsek model,
\beq
\tau^\mathrm{VM} \sim \frac{\eta}{J n_c a^2 k^2}  \ .
\label{naso}
\eeq
The divergence of this relaxation time for $k\to 0$ in the bulk is the dynamical side of
the divergence of the correlation length in the same model. 
At finite size there cannot be any real divergence; following the dynamic scaling hypothesis \eqref{hh1} 
and considering that in the scale-free case we have $\xi\sim L$, we can write,
\beq
\tau^\mathrm{VM}(k;L)  \sim \frac{1}{k^z} h(kL) \ ,
\label{oliviero}
\eeq
where $h$ is a scaling function. From \eqref{naso} we see that naive dimensional analysis gives for the Vicsek model, 
\beq
z^\mathrm{VM}_\mathrm{naive} =2 \ .
\eeq
However, as in the static case, the naive dynamical critical exponent can get corrections from renormalization and off-equilibrium effects,
so that its value could very well be different from $2$.

The second part of the dynamic scaling hypothesis, equation \eqref{hh2}, implies that the 
spatio-temporal correlation function depends on the size $L$ exclusively through the two factors
$t/\tau^\mathrm{VM}(k;L)$  and $kL$, hence giving,
\beq
C(k,t; L) = C_0(k;L)\;  f\left( \frac{t}{\tau^\mathrm{VM}(k;L)}, kL \right) \ .
\eeq
Using \eqref{oliviero}, we obtain,
\beq
C(k,t; L) =C_0(k;L)\;  f\left(\frac{k^zt}{h(kL)}, kL\right) \ .
\eeq
As first pointed out in  \cite{HH1967scaling}, we can now eliminate the dependence on the size $L$ by doing two things: first, we isolate the time-dependent part by defining the 
normalized dynamical correlation, i.e. the correlation divided by its static value for $t=0$,
\beq
\hat C(k,t; L) \equiv \frac{C(k,t;L)}{C_0(k;L)}  \ .
\eeq

\clearpage

\onecolumngrid

%%%%%%%%%%%%%%%%%%%%%%%%%%%%%%%%%%%%%%%%%%%%%%%%%%%
\begin{figure}[!h]
\centering
\includegraphics[width=0.7 \textwidth]{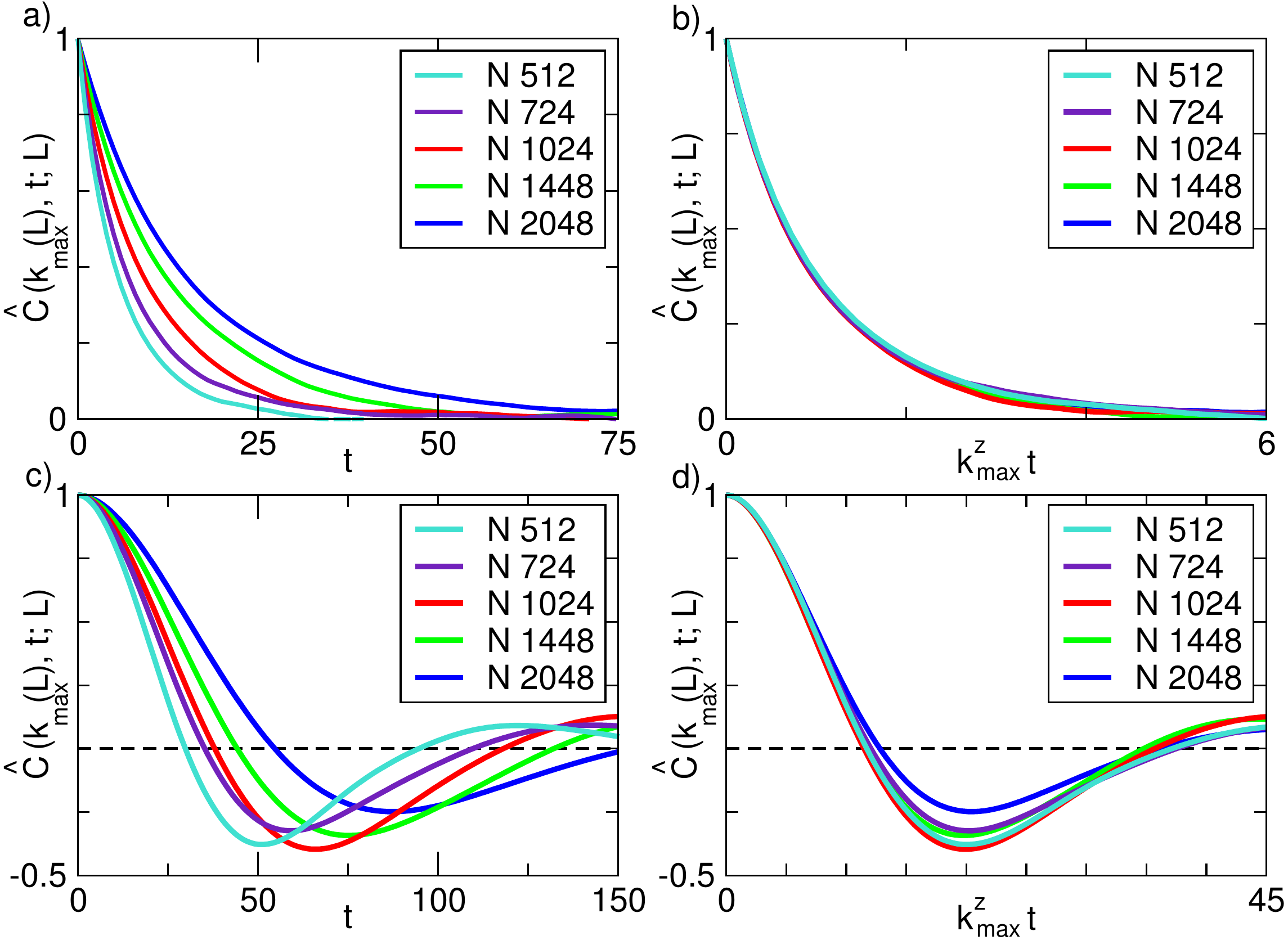}
\caption{{\bf Spatio-temporal correlation and scaling.} Numerical simulations in $d=3$ of the Vicsek model - panels (a) and (b) - and
of the inertial spin model - panels (c) and (d). a) Spatio-temporal correlation of the Vicsek model as a function of time at various sizes $L$ for $k=k_\mathrm{max}(L)$,
the maximum of the static correlation. b) Rescaled correlation of the Vicsek model; the best collapse occurs for $z=2.13$. c) Spatio-temporal correlation of 
the inertial spin model as a function of time at various sizes $L$ for $k=k_\mathrm{max}(L)$. d) Rescaled correlation of the inertial spin model; the best collapse occurs for $z=1.15$. }
\end{figure}
%%%%%%%%%%%%%%%%%%%%%%%%%%%%%%%%%%%%%%%%%%%%%%%%%%%
\twocolumngrid
\noindent 

\noindent
Secondly, we recall that the maximum of the static correlation occurs at 
$k_\mathrm{max} \sim 1/L$ (equations \eqref{ciupa} and \eqref{nemo}); therefore, if we evaluate the
normalized spatio-temporal correlation at this special momentum of maximal static correlation, we get,
\beq
\hat C(k_\mathrm{max}(L),t; L) = f\left(\frac{k_\mathrm{max}^zt}{h(1)},1\right) = \hat f\left( k_\mathrm{max}^zt \right) \ .
\label{sciato}
\eeq
This equation implies that it must exist a dynamical critical exponent, $z$, such that the spatio-temporal correlation functions calculated at different values of $L$ 
collapse onto the same $L$-independent master curve, provided that we evaluate each correlation at $k=k_\mathrm{max}(L)$ and plot them as a function of 
the scaling variable $k_\mathrm{max}^z t$. This collapse is the most conspicuous and easy-to-test prediction of the dynamic scaling hypothesis \cite{HH1967scaling}. 

Numerical simulations of the Vicsek model are in very good agreement with this prediction (Figs. 3a and 3b). 
Dynamical correlations at different sizes $L$ collapse rather well as a function of the scaling variable $k_\mathrm{max}^z t$. 
The best collapse is achieved for,
\beq
z^\mathrm{VM}= 2.13 \pm 0.02  \ ,
\label{pera}
\eeq
very close to the naive value of the dynamical critical exponent, indicating, as in the static case, 
that corrections to naive dimensional analysis exponents are somewhat weak. Notice that if we  keep $k$ fixed for different values of $L$, rather than following $k=k_\mathrm{max}(L)$, 
it becomes impossible to collapse the curves. As stated first in \cite{HH1967scaling}, following the peak of the static correlation by keeping $k\xi$ fixed is indeed necessary to make dynamic scaling work.

%%%%%%%%%%%%%%%%%%%%%%%%%
\subsection{Dynamical correlation: Inertial case}
%%%%%%%%%%%%%%%%%%%%%%%%%

In Section II naive dimensional analysis of the inertial dynamics of the ISM  provided two time scales,
\beq
 \tau_1^\mathrm{ISM} \sim 1/\gamma\quad , \quad \tau_2^\mathrm{ISM} \sim \frac{1}{c_sk}  \ .
\label{zonza}
\eeq
Hence, in the naive case $\tau_1^\mathrm{ISM}$ does not depend on $k$, whereas $\tau_2^\mathrm{ISM}$ does;  
it is therefore highly probable that the two time scales will have a different dependence on $k$ also in the general case.
This fact has an unpleasant consequence:  if we follow the dynamic scaling hypothesis and assume 
that the normalized spatio-temporal correlation function depends on $L$
only through its two time scales and through the factor $kL$, we obtain,
\beq
\hat C(k,t; L) = g\left( \frac{t}{\tau_1^\mathrm{ISM}(k,L)} , \frac{t}{\tau_2^\mathrm{ISM}(k; L)},kL \right) \ ,
\label{toro}
\eeq
from which it would seem impossible to scale {\it both} time scales and obtain a collapse of the correlation 
functions at different values of $L$ as we had in the non-inertial case.

Fortunately, there is a way out of this problem. We are interested in the underdamped regime of the inertial model, which corresponds to 
having a damping time much larger than the time of
propagation of mode $k$,
\beq
\tau_1^\mathrm{ISM}\gg \tau_2^\mathrm{ISM} \ .
\label{pelo}
\eeq
If we ask this condition to hold for all physical modes, including the smallest one, which is of order $1/L$, 
we get,
\beq
L/c_s \ll \tau_1^\mathrm{ISM} \ .
\label{sauro}
\eeq
whose meaning is clear: the time a signal takes to cross the system ($\sim L/c_s$) must be much shorter than 
the time the signal takes to get damped ($\sim\tau_1^\mathrm{ISM}$), which is a rather reasonable definition
of the underdamped phase. Notice that \eqref{sauro} can be rewritten as, 
\beq 
1/L \gg  \frac{\gamma}{c_s} \equiv k_0 \ .
\label{palo}
\eeq
In other words, there is a threshold momentum, $k_0$, which separates the overdamped phase ($1/L \ll k_0$), 
from the underdamped phase ($1/L \gg k_0$) \cite{cavagna+al_15}. We are interested in a system
where information can propagate, hence we consider the underdamped regime \eqref{pelo}-\eqref{palo}.
Moreover, we will consider short times, which is the regime in which it is the easiest to obtain the correlation 
function with good accuracy, especially in real experiments. Hence, we work in the following condition,
\beq
t \sim \tau_2^\mathrm{ISM} \ll \tau_1^\mathrm{ISM} \ ,
\eeq
under which the normalized correlation \eqref{toro} becomes a function of just {\it one} time scale, 
\beq 
\hat C(k,t; L) =  g\left(t/\tau_2^\mathrm{ISM} ,kL  \right) \ .
\label{sega}
\eeq
We can now use the same scaling procedure as in the non-inertial case.
We first extend the dynamic scaling hypothesis \eqref{hh1} to the inertial time scale $\tau_2^\mathrm{ISM}$, 
\beq
\tau_2^\mathrm{ISM}(k;L)  \sim \frac{1}{k^z} h(kL) \ ,
\label{toscani}
\eeq
where, again, we have left the dynamical exponent $z$ free to take a value different from 
that of naive dimensional analysis,
\beq
z^\mathrm{ISM}_\mathrm{naive}= 1 \ .
\eeq
We  then evaluate the dynamical correlation at the maximum of the static correlation, $k=k_\mathrm{max}\sim 1/L$,
which gives,
\beq
\hat C(k_\mathrm{max}(L),t; L) \sim  g\left(\frac{k_\mathrm{max}^zt}{h(1)}, 1 \right) = \hat g\left(k_\mathrm{max}^zt \right)  \ .
\label{sciuto}
\eeq
To check this prediction of dynamic scaling we performed three-dimensional simulations of the inertial spin model, Figs. 3c and 3d. Results confirm fully the validity of the scaling equation \eqref{sciuto}.  
We notice that the correlation function has quite a different form from the non-inertial one: there
are clear oscillations, a clear fingerprint of the underdamped regime of the inertial dynamics \cite{cavagna+al_15}; moreover, the correlation
is quadratic, rather than linear, in the limit $t\to 0$. Yet the most clearcut difference with the non-inertial case is provided by the different
value of the dynamical exponent; the  best collapse of the data is given by, 
\beq
z^\mathrm{ISM} = 1.15 \pm 0.02 \ ,
\label{nobu}
\eeq
well distinguishable from the Vicsek value, relation \eqref{pera}. Again, we find an exponent very close to 
its naive counterpart, indicating that off-equilibrium and renormalization corrections are weak.

%%%%%%%%%%%%%%%%%%%%%%%%%%%%%%%%%%%%
\begin{figure}[!h]
\centering
\includegraphics[width=0.5 \textwidth]{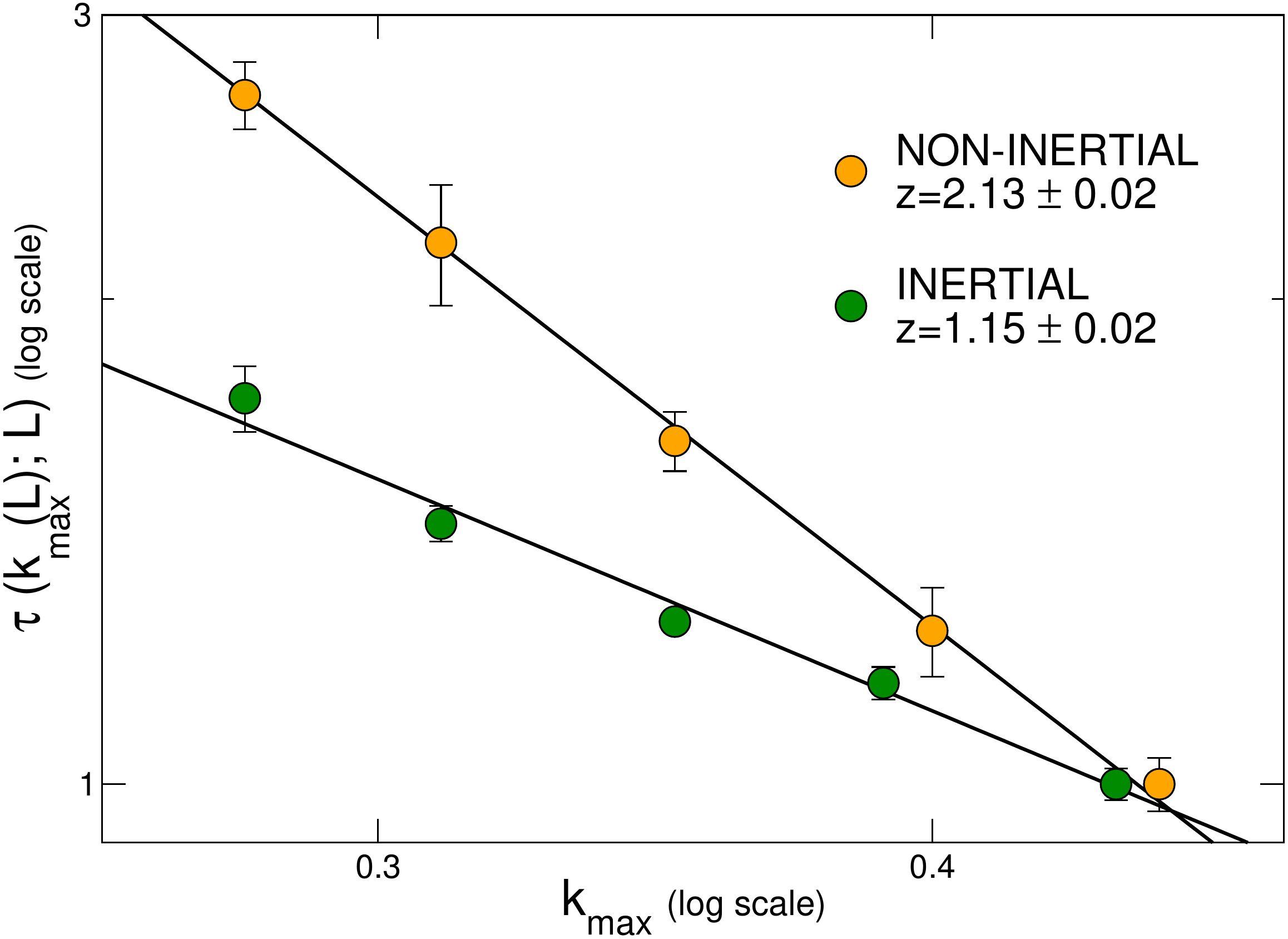}
\caption{{\bf Times scales of the two models}. We computed the time scales of the Vicsek model (orange points) and of the inertial spin model
(green points) by crossing the spatio-temporal correlations in Fig. 3 with a fixed value, $\hat C(k_\mathrm{max}, \tau)= 1/e$, then we have normalized the values obtained in such a way that in each of the two cases the smallest $\tau$ was equal to $1$.
The dependence of the time scales with the momentum is rather different in the two models, due to the difference in the two critical dynamical exponents.
}
\end{figure}
%%%%%%%%%%%%%%%%%%%%%%%%%%%%%%%%%%%%%%%%%%%%%%%%%

To conclude we test equations \eqref{oliviero} and \eqref{toscani} about the scaling behavior with $k$ of the  
characteristic time scales in the two different models. As usual, to eliminate the dependence on the
scaling function, $h(kL)$, we need to work at the maximum of the static correlation, $k=k_\mathrm{max}\sim 1/L$. 
To extract the time scale $\tau$ from Fig. 3a and 3c we use the simple crossing condition,
\beq
\hat C(k_\mathrm{max}, \tau)= 1/e  \ .
\eeq
In Fig. 4 we report $\tau(k_\mathrm{max})$ vs $k_\mathrm{max}$ at various values of the size $L$ in the non-inertial case (orange points) and
in the inertial case (green points). As predicted by equations \eqref{oliviero} and \eqref{toscani}, the characteristic time scale depends on 
the momentum as a power law. Moreover, the critical dynamical exponents, $z$, obtained by the dynamic scaling, equations \eqref{pera} and \eqref{nobu}, fit rather well the data
and differentiate sharply the two cases.

Note that, although the two time scales have been obtained by using exactly the same protocol in the two models 
 - calculating the correlation $C(k,t)$, evaluating it at $k_\mathrm{max}(L)$, crossing the correlation with a constant -  they have 
a rather different physical meaning: in the non-inertial case $\tau$ is a damping time of an overdamped correlation function, while in the inertial case $\tau$ is
actually the period of an underdamped, oscillating correlation function. 

The results of this Section fully demonstrate that a finite-size scaling analysis of the spatio-temporal correlation function successfully 
distinguishes between the non-inertial dynamics of the Vicsek model and the inertial dynamics of the ISM. The fundamental fact 
underlying this result is that the intrinsic, merely dimensional, time scales of the two systems depend on the momentum $k$ in such 
sharply different ways that off-equilibrium
corrections are unable to wash out this distinction, at least in the ordered phase we are considering. 
In the next Section we will compute explicitly the dispersion relations of 
the two models under an approximate scheme, hence making even more clear the different mathematical structure of inertial and
non-inertial models. In the present Section, however, we stress that we have made no particular approximation in the derivation of the 
finite-size scaling relations. The only simplification that we have adopted has been to neglect the difference between 
longitudinal and transverse direction (and therefore momentum) and to describe everything as a function of the scalar momentum $k$.
Our exact numerical simulations of the actual off-equilibrium, self-propelled models fully confirm the theoretical expectations 
of dynamic scaling.

%%%%%%%%%%%%%%%%%%%%%%%%%%%%%%%%%%%%%%%%
\subsection{Summary of the method}
%%%%%%%%%%%%%%%%%%%%%%%%%%%%%%%%%%%%%%%%

At this point our procedure may seem rather intricate. In fact, it is not; it simply consists of 
three steps,  let us briefly summarize them here.

\subsubsection{Calculate $C(k,t;L)$ from the data}

Given a certain dataset, the first thing to do is to compute the spatio-temporal correlation function in Fourier space, $C(k,t;L)$, 
using definition \eqref{dangle}. Clearly, for the method to work it is vital to have
data at different sizes $L$. We  recall that to compute the correlation function  one must perform
a time average (average over $t_0$). As always when computing time-correlation functions, the {\it total} time of the simulation 
(or of the experiment) must be much larger than the relaxation time.
In experiments, however, this is not always possible. Some help comes from the fact that our scaling relations are particularly strong
for short times, which are more experimentally accessible. However, the time averages must be at least long enough to stabilize 
the static correlation, $C_0(k;L)$.

Regarding the interval in $k$ to be considered when computing $C(k,t)$, we notice that the natural upper limit is the inverse 
of the mean interparticle distance, $1/a$. Well above this point the $\sin(kr_{ij})/kr_{ij}$ factor oscillates very strongly. 
On the other hand, although a natural scale for the minima value of $k$ is $1/L$, the correlation function 
must be calculated down to $k=0$, in order to check that $C(k=0,t)=0$ and to clearly see the maximum of the 
static correlation.

\subsubsection{Find the peak of the static correlation $C_0(k;L)$}

Once the full spatio-temporal correlation, $C(k,t; L)$, is calculated, one must plot its static limit, namely its amplitude, $C_0(k;L)=C(k, t=0;L)$, as a function of $k$. 
This function must be zero at $k=0$ and (as long as the system has non-negligible correlation length) it has a maximum at some intermediate $k_\mathrm{max}$ (Figs. 1 and 2). This maximal momentum corresponds to the inverse correlation length, $k_\mathrm{max}\sim 1/\xi$, and in a scale-free system 
it will scale as $k_\mathrm{max}(L)\sim 1/L$. On the other hand, the value of the static correlation at $k_\mathrm{max}$ is the best estimate of 
the susceptibility and it scales as some power of the size, $C_0(k_\mathrm{max})\sim L^{\gamma/\nu}$. Both these relations should be checked
for consistency.

\subsubsection{Collapse the dynamical correlations at different sizes}

For each size $L$, one must evaluate the normalized spatio-temporal correlation function $\hat C(k,t;L)$ at $k=k_\mathrm{max}(L)$. All these curves must be 
plotted against the rescaled time, $k_\mathrm{max}^z t$, and one must find the value of the dynamical exponent $z$ that 
produces the best collapse of all the curves at different sizes $L$ (Fig. 3).
The value of the dynamical exponent $z$ can then be compared in different models to distinguish their dynamics. Our
analysis shows that, {\it in the ordered phase}, a large exponent, $z\sim 2$, is associated to the non-inertial, overdamped dynamics of the Vicsek type, 
while a small exponent, $z\sim 1$, is associated to the inertial, underdamped dynamics of the ISM type. 
As we wrote in the Introduction, an equivalent analysis 
close to the ordering transition should be conducted to see whether or not a similar change in the dynamical exponent occurs in  `swarm-like' 
collective systems.

%%%%%%%%%%%%%%%%%%%%%%%%%%%%%%%%%%%%%%%%%%%%%
\section{Approximate theory}
%%%%%%%%%%%%%%%%%%%%%%%%%%%%%%%%%%%%%%%%%%%%%

In this Section we perform an analytical calculation of the spatio-temporal correlation function, $C(k,t)$, based on an approximate scheme.
As we shall see, the results are in line with the general theory describe above.

%%%%%%%%%%%%%%%%%%%%%%%%
\subsection{Fixed network approximation}
%%%%%%%%%%%%%%%%%%%%%%%

The first approximation we adopt is that of fixed network. This approximation has the great advantage of enormously simplifying the computations, 
but it may seem rather extreme given the very nature of active systems. 
We briefly discuss here the nature, limitations and range of applicability of this approximation. For a more detailed discussion we refer to \cite{mora2015questioning}.

Biological systems displaying collective motion  differ from traditional physical systems because they are inherently out of equilibrium: its constituents are particles that move by self-propulsion, constantly compensating for the dissipation effects by injecting energy into the system. The key ingredient of an active system is the rearrangement of the interaction network, a phenomenon that has very important effects \cite{toner+al_95,toner+al_98,toner1998flocks}.

Taking into account the full active nature of these systems at a theoretical level requires a hydrodynamic approach \cite{toner+al_95,toner+al_98,toner1998flocks, ramaswamy_review}, which describes the coupling between velocity and density fields, providing an elegant description of the large scale behaviour of active fluids. In order to select 
the relevant terms in the continuum equations, this approach focuses on large length scales, i.e. on the so-called hydrodynamic limit $k\to 0$. Inertial effects, however, are completely dominated by dissipation in this limit. And yet we know as an experimental fact that inertial effects are 
essential to reproduce real flock phenomenology: real flocks are finite-size systems, therefore far from the $k\to 0$ hydrodynamic limit, and for this reason inertial effects 
dominate over dissipation, not the other way around. The inevitable conclusion is that in order to describe real, finite-size biological groups, we need to give up the simplifying 
framework of the hydrodynamic limit. Even though this can still be done in the framework of a continuum theory of the velocity and density fields (see \cite{cavagna2015silent}), we
take here a different route.

Let us assume that, even though the particles of our system move, the rearrangement of the neighbours' interaction network is {\it slow}. By this we mean that there is a separation of timescales: if we consider the scale of local relaxation, $\tau_{\mathrm{relax}}$, defined as the characteristic time needed to relax locally the order parameter with the interaction network fixed, and the network reshuffling time, $\tau_{\mathrm{network}}$, that is the average time it takes for an individual to change its interacting neighborhood, then by slow network rearrangement we mean \cite{mora2015questioning},
\beq 
\tau_{\mathrm{relax}}\ll \tau_{\mathrm{network}}  \ .
\label{slow}
\eeq
Under this condition of local quasi-equilibrium, the update of the velocity of a particle occurs on a time-scale much faster than that needed to 
change the matrix of its neighbours, $n_{ij}(t)$, which can then be considered constant in time,
\beq
n_{ij}(t)\sim  n_{ij}  \ .
\label{guzo}
\eeq
Natural flocks of birds are precisely in such a state of local quasi-equilibrium \cite{mora2015questioning} and indeed the predictions for the propagation law based on the fixed network analysis work very well \cite{attanasi+al_14,cavagna+al_15}. Of course there must be a crossover length scale $l^\star$ beyond which the network rearrangements become relevant and a hydrodynamic approach is mandatory. The fixed network approximation therefore describes all the modes with $k > k^\star= 1/l^\star$. However, as we stated above, experimental evidence shows that real flocks are well within this scale, $L< l^\star$, and this is why we adopt approximation \eqref{guzo}.

%%%%%%%%%%%%%%%%%%%%%%%%%%%%%%%%%%%%
\subsection{Spin-wave expansion and continuous limit}
%%%%%%%%%%%%%%%%%%%%%%%%%%%%%%%%%%%%

In this work we are considering systems in their strongly polarized phase. To fix ideas, we shall assume that the mean velocity 
of the group is pointing in direction $x$, i.e. along the unit vector ${\bf n}_x = (1,0,0)$.
Each velocity ${\bf v}_i$ can be decomposed into a longitudinal component, let us call it $v_i^x $, along the direction of motion 
${\bf n}_x$ and a transverse component, which is a $(d-1)$-dimensional vector $\boldsymbol{\pi}_i$ lying on the plane 
perpendicular to the direction of motion, 
\begin{equation}
{\bf v}_i = v_i^x {\bf n}_x + \boldsymbol{\pi}_i \ .
\end{equation}
Notice that the transverse components $\boldsymbol{\pi}_i$ have the physical dimension of a velocity and they satisfy the obvious relation, 
\beq
\sum_i \boldsymbol{\pi}_i = 0 \ .
\eeq
Given that we are studying models with fixed speed, $|{\bf v}_i|=v_0$, we can work out the longitudinal component as a function of
the transverse one,
\beq
v_i^x = \sqrt{v_0^2 - \pi_i^2}  \ .
\eeq
When the polarization is large all velocities will be mainly along the mean direction of motion, 
implying $\pi_i^2 \ll v_0$. This is the so-called {\it spin-wave approximation}, which yields,
\begin{equation}
v_i^x  \sim v_0 \left(1 - \frac{1}{2}\pi_i^2 /v_0^2\right) \ ,
\end{equation}
and,
\begin{equation}
{\bf v}_i =  {\bf n}_x \, v_0 \, \left(1 - \frac{1}{2}\pi_i^2 /v_0^2\right) + \boldsymbol{\pi}_i  \ .
\label{nando}
\end{equation}
% Phases
It is convenient to write the transverse components of the velocity, $\boldsymbol{\pi}_i$, in terms of dimensionless angles 
expressing the departure of each ${\bf v}_i$ from the mean direction of motion, ${\bf n}_x$,
\bea
\pi_i^y &&= v_0 \sin \varphi_i^z \sim v_0 \, \varphi_i^z \ ,
\label{nandu}
\\
\pi_i^z &&= v_0 \sin \varphi_i^y \sim v_0 \, \varphi_i^y  \ .
\label{nanda}
\eea
To understand these relations we must recall that to create a $y$ component of the velocity one needs to rotate ${\bf v}_i$ around the $z$ axis, and vice-versa. These transverse angles $\varphi_i^z$ and $\varphi_i^y$ are the key degrees of freedom in a polarized system and they are 
called {\it phases}. They simply represent the (small) angular deviations of each individual ${\bf v}_i$ with respect to the mean velocity of the group. 

% Spin wave expansion delle equazioni dinamiche
We can now plug  equations \eqref{nando}, \eqref{nandu} and \eqref{nanda} into the full dynamical equations describing the models and expand them 
up to the first order in the phase, so to obtain equations directly for the $\varphi_i$. This is called {\it spin-wave} expansion \cite{dyson_56}.
For the Vicsek model \eqref{vic1} the spin-wave expansion gives the same equation  for both $\varphi^y$ and $\varphi^z$, namely, 
\begin{equation}
\eta \frac{d \varphi_i}{dt}= - J \sum_j \Lambda_{ij} \varphi_j + \zeta_i^{\perp}  \ ,
\label{swvic}
\end{equation}
where $\Lambda_{ij}$ is the Laplacian matrix defined in \eqref{laplacian}. Similarly expanding the ISM equation \eqref{grand}, we obtain
\begin{equation}
\chi \frac{d^2 \varphi_i}{dt^2} + \eta \frac{d \varphi_i}{dt}= - J \sum_j \Lambda_{ij} \varphi_j + \zeta_i^{\perp}  \ .
\label{swism}
\end{equation}
% Polarization
From relations \eqref{nando}, \eqref{nandu} and \eqref{nanda} we can also work out an expression of the polarization $\Phi$ in terms of the phase, 
\beq
\Phi \equiv  \left| \frac{1}{N}\sum_i \frac{{\bf v}_i}{v_0} \right|= 1 - \frac{(d-1)}{2N}\sum_i  \varphi_i^2 \ ,
\label{susa}
\eeq
from which we see that the limit of large polarization, $\Phi \sim 1$, is equivalent to the limit of small phases,
$\varphi_i^2 \ll 1$.

If we look at spatial scales larger than the nearest neighbor distances -- and we must do that, lest all of our scaling relations lose their validity-- we can approximate the discrete Laplacian with its continuous counterpart (we recall that $a$ is the mean interparticle distance),
\beq
J \sum_j \Lambda_{ij} \rightarrow - J n_c a^2 \nabla^2  \ .
\eeq
In performing this substitution it is of course crucial the previous fixed-network assumption, that is the fact that $\Lambda_{ij}$ does 
not depend on time.
Similarly, we can substitute the discrete-space phases with continuous fields, 
\beq
\varphi_i(t) \to \varphi({\bf x},t) \ .
\eeq
In this way we can rewrite the Vicsek model as, 
\begin{equation}
\left(\eta \frac{\partial }{\partial t} - J n_c a^2 \, \nabla^2 \right)\varphi({\bf x}, t) =\zeta({\bf x}, t) \ .
\label{zombie}
\end{equation}
whereas for the inertial spin model, we obtain,
\begin{equation}
\left( 
\chi \frac{\partial^2}{\partial t^2}  + \eta \frac{\partial}{\partial t} - J n_c a^2\, \nabla^2 
\right) \varphi({\bf x}, t) = \zeta({\bf x}, t)  \ .
\label{ismsw}
\end{equation}
In both cases $\zeta$ is a Gaussian white noise, 
\beq
\langle \zeta({\bf x}, t) \zeta ({\bf x}', t') \rangle= 2  \eta \, T \, a^3\delta^{(3)}({\bf x}- {\bf x}')\delta(t-t') .
\eeq
where the factor  $a^3$ is necessary to keep the original physical dimensions once we introduce the spatial Dirac's delta.

% Fluttuazioni di velocita' in spin-wave
In the highly polarized phase, from the definition of velocity fluctuations \eqref{fluctu}, and from \eqref{nandu} and \eqref{nanda}, we obtain,
\beq 
\delta \hat{\bf v}_i  = \frac{1}{\sqrt{1 - \Phi}} (0,\ \varphi_i^z,\  \varphi_i^y)  \ ,
\label{uppa}
\eeq
up to linear order in the phase. This equation embodies the fact that in a polarized system the fluctuations are strongly dominated by 
their transverse components. Accordingly, the connected velocity correlation \eqref{minkia} becomes,
\beq
C_{ij}=  \langle \varphi_i \, \varphi_j \rangle \ ,
\eeq
%\end{widetext}
up to a constant factor equal to $(d-1)/(1-\Phi)$, and where we have taken into account the fact that the statistical correlation of $\varphi^y$ is the same as that of $\varphi^z$, as they satisfy identical stochastic equations. 
In the light of this result the spatio-temporal correlation function, $C({\bf r},t)$, defined in \eqref{mingus} 
can be written as,
\beq
C({\bf r},t) =  \left\langle \frac{1}{V} \int d{\bf x}_0\  \varphi({\bf x}_0,t_0) \varphi({\bf x}_0+{\bf r},t_0+t) \right\rangle_{t_0} \ ,
\label{cicuta}
\eeq
%\end{widetext}
which in $k$-space becomes, 
\beq
C({\bf k}, t) =\langle \varphi({\bf k}, t_0)\varphi(-{\bf k }, t_0+t)\rangle \ .
\label{topeko}
\eeq
This compact form of the correlation will be particularly useful for the theoretical calculations of the next Section.
\medskip

%%%%%%%%%%%%%%%%%%%%%%%%%%%%%%%%%%%
\subsection{Theory: Non-inertial dynamics}
%%%%%%%%%%%%%%%%%%%%%%%%%%%%%%%%%%%

The linear stochastic differential equations \eqref{zombie} and \eqref{ismsw} can be easily solved by using the Green functions method \cite{lanczos1961linear}, whose details are described in Appendix B. The Green function, $G({\bf k}, \omega)$, 
is essentially the inverse, in Fourier space, of the differential operator ruling a dynamical equation. For the Vicsek model, we have from \eqref{zombie}, 
\begin{equation}
G({\bf k}, \omega) = \frac{1}{ i \eta \omega + J n_c a^2 k^2}  \ .
\label{grongo}
\end{equation}
Once the Green function is known, the correlation function \eqref{topeko} is given by,
\beq
C({\bf k}, t) =\frac{2 \eta T\,a^3}{2 \pi} \int d\omega \,  \, e^{i \omega t} \ G({\bf k}, \omega) \, G(-{\bf k}, -\omega)  \ .
\label{zut}
\eeq

\clearpage

\onecolumngrid

%%%%%%%%%%%%%%%%%%%%%%%%%%%%%
\begin{figure}[!h]
\centering
\includegraphics[width=0.7\textwidth]{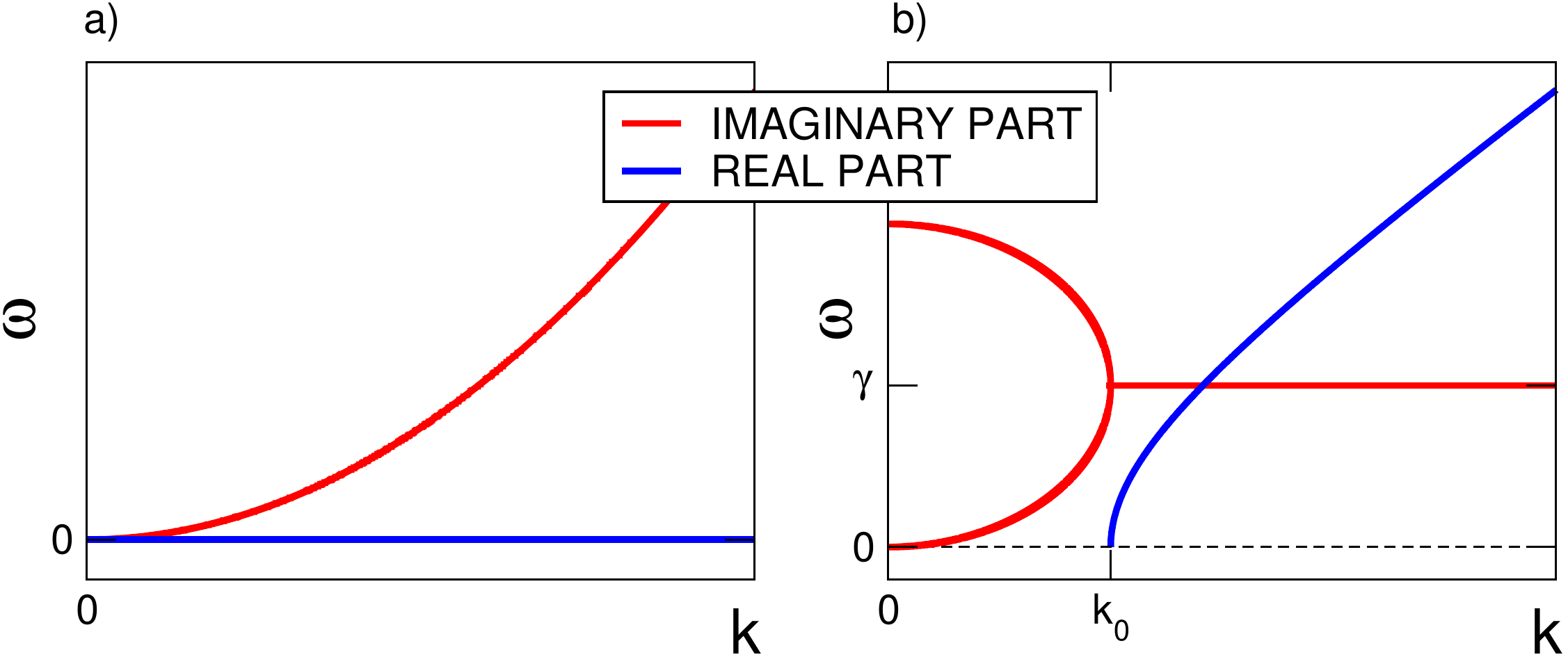}
\caption{{\bf Dispersion relations.} Real (blue) and imaginary (red) part of the frequency $\omega$ as a function of the momentum $k$.
{\bf a)} -- Vicsek model: the frequency is purely imaginary, as the system is overdamped and no signal propagation occurs.
{\bf b)} -- Inertial spin model: for $k>k_0$ the frequency develops a real part, giving rise to signal propagation. Notice that for $k\to\infty$ the imaginary part 
grows asymptotically as $k^2$ in the non-inertial case, while it saturates to $\gamma$ in the inertial case. On the other hand, for $k\to 0$, the inertial 
and non-inertial frequencies coincide: both real parts are zero and both imaginary parts go to zero as $k^2$.
}
\end{figure}
%%%%%%%%%%%%%%%%%%%%%%%%%%%%%%%%
\twocolumngrid

\noindent The frequency integral is performed by Cauchy's residue method, which consists in evaluating the integrand at its simple poles in the complex $\omega$ plane. 
For this reason the frequencies at which the Green function has the poles acquire particular importance; these frequencies are defined by the so-called {\it dispersion relation}, which in this case of Vicsek non-inertial dynamics reads,
\beq 
i \eta \omega+ J n_c a^2 k^2=0 \ .
\label{fiori}
\eeq
The frequency $\omega$ is purely imaginary with a quadratic (i.e. diffusive) dispersion law,
\begin{equation}
\omega = i \; \frac{J n_c a^2 }{\eta}k^2 \ .
\label{tony}
\end{equation}
The integral in \eqref{zut} can be easily performed, giving the dynamical correlation function of the Vicsek case,
\begin{equation}
C^\mathrm{VM}(k,t)= C_0(k) \, e^{- \frac{J n_ca^2}{\eta} k^2 t }  \ .
\label{viccorr}
\end{equation}
where the static correlation function is given by, 
\beq
C_0(k) =\frac{2 T a}{J n_c k^2}\ .
\label{piro}
\eeq
In the non-inertial Vicsek model the correlation function \eqref{viccorr} is therefore a pure exponential,
with relaxation time given by,
\beq
\tau^\mathrm{VM}(k) = \frac{\eta}{J n_c a^2 k^2}  \ ,
\label{zonzo}
\eeq
which is the same result that we obtained with naive dimensional analysis, equation \eqref{zappo}.

%%%%%%%%%%%%%%%%%%%%%%%%%%%%%%%%%%%%%%%%%
\subsection{Theory: Inertial dynamics}
%%%%%%%%%%%%%%%%%%%%%%%%%%%%%%%%%%%%%%%%%

The dynamics of the inertial spin model is given by equation \eqref{ismsw}, which gives the Green function,
\begin{equation}
G({\bf k}, \omega) = \frac{1}{-\chi \omega^2 + i \eta \omega + J n_c a^2 k^2} \ ,
\label{napoli}
\end{equation}
with dispersion law,
\begin{equation}
-\chi \omega^2 + i \eta \omega + J n_c a^2 k^2=0 \ .
\label{minni}
\end{equation}
This equation has two complex solutions,
\begin{equation}
\omega = i \frac{\eta}{2\chi} \pm \frac{1}{2\chi} \sqrt{4\chi J n_c a^2 k^2 -\eta^2  } \ .
\end{equation}
The first thing to notice is that in the $k\to 0$ limit we recover exactly the same dispersion law 
as in the non-inertial theory. Indeed, in this limit (that is in the hydrodynamic limit) we 
obtain two purely imaginary frequencies, the smallest of which is,
\beq
\omega(k) \sim  i \ \frac{J n_c a^2}{\eta} k^2 \quad , \quad k\to 0 \ ,
\eeq
equal to equation \eqref{tony}. This fact is a further confirmation that the inertial spin model gives in the hydrodynamic limit $k\to 0$ the same results as the Vicsek model \cite{cavagna+al_15}.

For generic $k$ the dispersion relation can be simplified by introducing the reduced friction coefficient, $\gamma$, and the second sound speed, $c_s$, 
previously defined in \eqref{zappa}, and the threshold momentum, $k_0$, defined in \eqref{palo}. In this way we obtain,
\begin{equation}
\omega = i \gamma \pm c_s k \ \sqrt{1 - k_0^2/k^2} \ ,
\label{frigo}
\end{equation}
We see that for $k > k_0$ the frequency has nonzero real part, so that there is signal propagation,
while for $k<k_0$ the frequency is purely imaginary and the dynamics is overdamped, as in the Vicsek case.
In the deeply underdamped regime, $k\gg k_0$, the dispersion relation further simplifies and we get,
\beq
\omega = i\gamma \pm c_s k \ .
\eeq
In this case each mode $k$ propagates linearly with the same speed, $c_s$, and damping $\gamma$. 
The two different dispersion relations, \eqref{tony} and \eqref{frigo} are depicted in Fig.5.

By plugging the inertial Green function \eqref{napoli} into \eqref{zut} and performing the residue integral in the complex $\omega$ plane, 
we obtain the spatio-temporal correlation function in the inertial case,
\begin{widetext}
\begin{equation}
C^\mathrm{ISM}(k,t)= C_0(k) \ 
e^{- \gamma t } \, 
\left[  
 \frac{\gamma}{c_s k } \frac{1}{\sqrt{1-k_0^2/k^2}}      \sin{\left( c_s k \, t \; \sqrt{ 1- k_0^2/k^2} \right)}+ 
\cos{\left( c_s k \, t \; \sqrt{ 1- k_0^2/k^2} \right)} \right] \ ,
\label{piropiro}
\end{equation}

\end{widetext}
where that the static correlation function, $C_0(k)$ is the same as in the non-inertial case, equation \eqref{piro}:
as we already remarked several times, the static correlation function does not distinguish between different dynamics.
In the deeply underdamped regime, $k \gg k_0$, the inertial correlation function takes the simpler form,
\beq
C^\mathrm{ISM}(k,t)= C_0(k) \ e^{- \gamma t } \, \left[   \frac{\gamma}{c_s k }  \sin{\left( c_s k \, t \right)}+ 
\cos{\left( c_s k \, t \right)} \right] \, ,
\eeq
This spatio-temporal correlation is completely different from the non-inertial case: it is an oscillating function of time, characterized by two time scales,
\beq
\tau_1^\mathrm{ISM} = 1/\gamma, \quad\quad  \tau_2^\mathrm{ISM} = \frac{1}{c_s k}
\eeq
Again, we recover the same time scales as in the naive dimensional analysis, relations \eqref{zappe}.
The expansion for short times of the normalized correlation function in the inertial case, equation \eqref{piropiro}, gives,
\beq
\hat{C}^\mathrm{ISM}(k,t) \sim 1 - \frac{1}{2} (c_s kt)^2 \, .
\eeq
This result shows that for short times: (i) the inertial correlation function decays quadratically, unlike the linear decay of the non-inertial case; (ii) the inertial correlation function depends on just one time scale, namely the period,
\beq
\hat{C}^\mathrm{ISM}(k,t) \sim 1 - \frac{1}{2} (t/\tau_2^\mathrm{ISM}(k))^2 \, .
\eeq
which is the result we anticipated in \eqref{sega}, on which is based the finite-size scaling analysis of the inertial case.

The approximate scheme we presented (fixed network, large polarization and continuous limit) gives rise to a linear theory
which is essentially the massless Gaussian field theory \cite{binney_book}. From equations \eqref{piro}, \eqref{zonzo} and \eqref{zonzak} we see  
the critical exponent are given by,
\beq
(\gamma/\nu)_\mathrm{gauss} = 2 \quad , \quad z_\mathrm{gauss}^\mathrm{VM} = 2 \quad , \quad z_\mathrm{gauss}^\mathrm{ISM} = 1
\eeq
As usual in critical phenomena, the Gaussian approximation gives the same critical exponents as naive dimensional analysis \cite{binney_book}.
We have seen in the previous Sections that numerical simulations give critical exponents very close to the Gaussian/naive ones.

%%%%%%%%%%%%%%%%%%%%%%%%%%%%%%%%%%%
\subsection{The boundary of the approximate theory}
%%%%%%%%%%%%%%%%%%%%%%%%%%%%%%%%%%%

We have seen that in the context of the approximate theory we can work out an exact expression
for the polarization, equation \eqref{susa}. In $d=3$ we can rewrite that equation as,
\beq
\Phi = 1- \langle \varphi({\bf x})^2\rangle  \ ,
\eeq
where we have indicated with a bracket the space average. From \eqref{cicuta} we see that the 
average of the phase squared is simply the static spatio-temporal correlation function evaluated at ${\bf r}=0$,
namely,
\beq
\langle \varphi({\bf x})^2\rangle = C({\bf r}=0,t=0) = \int^{1/a} d{\bf k} \ C_0({\bf k}) \ ,
\eeq
where the upper limit of integration keeps into account the discrete nature of the system.
From equation \eqref{piro}, we finally obtain an explicit expression for the polarization as a function of the parameters of the model,
\beq
\Phi =  1 - \int^{1/a} d{\bf k} \ \frac{2 T a}{J n_c k^2}  = 1 - \frac{8 \pi T}{Jn_c} \ .
\label{buda}
\eeq
We tested this relation against numerical simulations. The results (which, being static, are identical for the Vicsek model and for the ISM)
are shown in Fig.6.

%%%%%%%%%%%%%%%%%%%%%%%%%%%%%%%%%%%%%%%%%
\begin{figure}[!h]
\centering
\includegraphics[width=0.4 \textwidth]{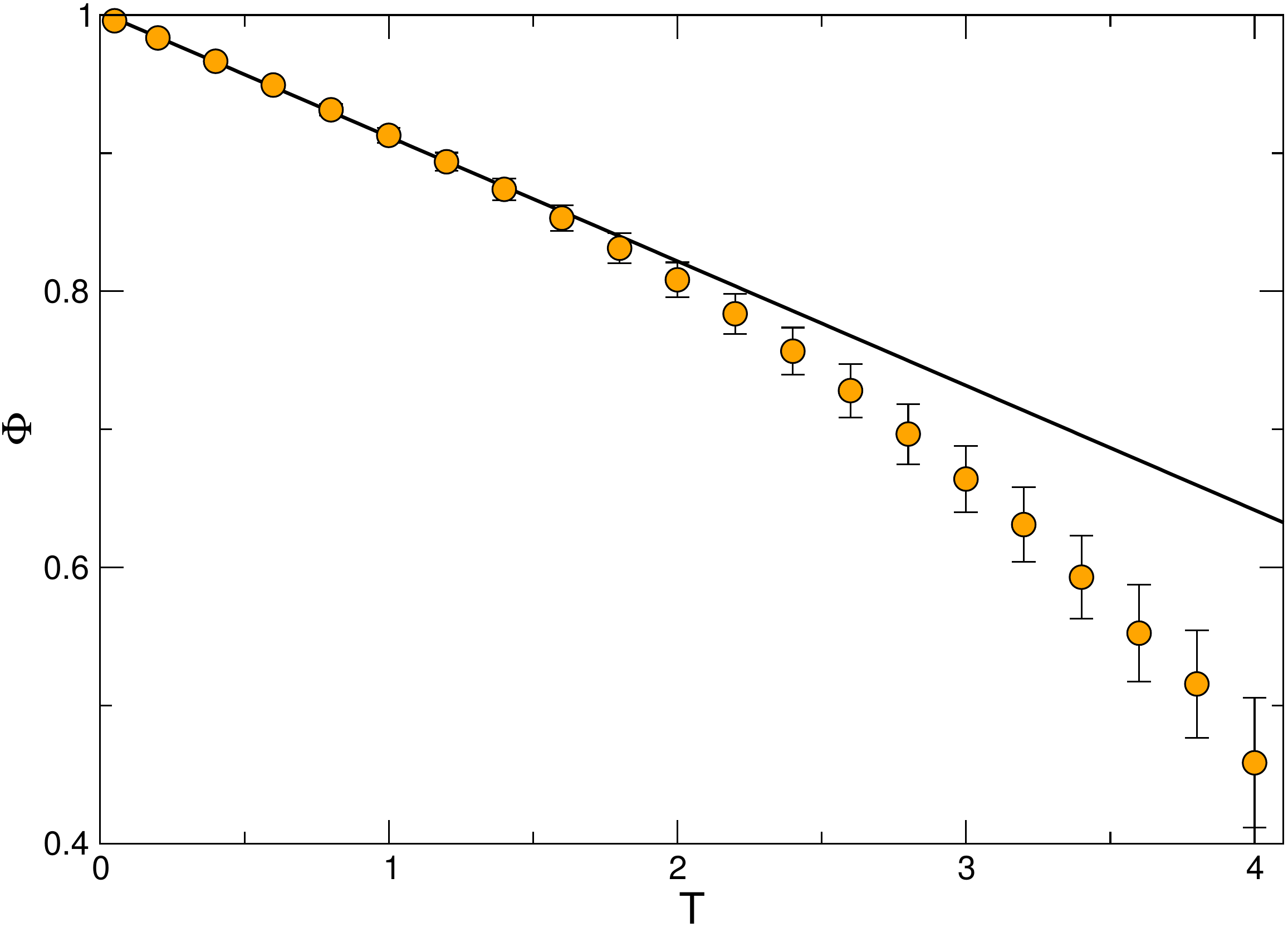}
\caption{{\bf Polarization.} We report here the polarization in the topological Vicsek model in $d=3$ as a
function of the temperature (results for the ISM and for metric interaction are identical). The full line is the spin-wave 
prediction \eqref{buda}. 
}
\end{figure}
%%%%%%%%%%%%%%%%%%%%%%%%%%%%%%%%%%%%%%%%%%

\noindent We remark that expression \eqref{buda} is valid only within the approximate scheme adopted
in this Section, namely: fixed network, large polarization (spin wave) and continuous limit. We have already seen that the critical exponents provided by
this approximation agree with those of exact numerical simulations, which were run at $\Phi = 0.9$. 
One may wonder up to what values of polarizations this will happen.
Fig.6 provides an answer: the spin-wave expansion (i.e. the Gaussian model) of the polarization is valid down to $\Phi \sim 0.7-0.8$.
Below these values the nonlinear corrections become significant. 
We notice that these values of the polarization are not outrageous: several biological groups display polarization larger than this value, 
which is therefore well within the range of the spin-wave expansion. 

Finally, we stress that the validity of \eqref{buda} implies that all the approximations used in this section are valid; on the contrary, the
break down of \eqref{buda} does not indicate which one of the approximations - fixed network, spin-wave, continuous limit - breaks down.

%%%%%%%%%%%%%%%%%%% end of theory %%%%%%%%%%%%%%%%%%%%%%%%%%%%%%%%%%%%

%%%%%%%%%%%%%%%%%%%%%%%%%%%%
\section{Conclusions}
%%%%%%%%%%%%%%%%%%%%%%%%%%%%

We have discussed a method able to distinguish models of collective motion with different dynamical behaviour. The method 
uses spontaneous fluctuations rather than explicit signal propagation across the system, which is very convenient, especially at
the experimental level. The key quantity of the method is the spatio-temporal correlation
function in Fourier space, $C(k,t)$, which we have defined in \eqref{dangle} in practical terms, easy to implement numerically and 
experimentally. 

We emphasize that the use of space-time correlations to infer information about the dynamics - more precisely,
to work out the dispersion law - is definitely not new. This is a standard procedure in equilibrium statistical physics, and it has
also been used in the  context of self-propelled particles models by Tu, Toner and Ulm \cite{toner+al_98}, who made a numerical
study of first-sound dispersion law in the Vicsek model. Our new contribution here has been to apply the method to inertial dynamics,
which had never been made before, and to use it as a tool to distinguish non-inertial from inertial dynamics.

Our simulations indicate that the method is very promising and that it may now be exported to experimental data on real biological systems.
We have, however, to be careful. As we have seen, in the limit $k\to 0$ all models (inertial and non-inertial), give the same result, namely 
the static correlation. Hence, the method we have described is fruitful at non-zero values of the momentum $k$. What does this mean at 
the quantitative level? The answer depends on the correlation length, $\xi$. We have seen that the crucial scale for the analysis is 
$k_\mathrm{max} \sim 1/\xi$. If the system is strongly
correlated, then $\xi$ will be large, so that the method will use information integrated over a large spatial scale (that is, summed over many individuals), 
thus providing an accurate signal. 
If, on the other hand, the system is poorly correlated, then $\xi$ is small, and the method integrates information on short
spatial scales; this is a problem, because short scales are much more prone to experimental error (mainly, but not solely, due to segmentation errors in 
the image analysis \cite{Cavagna2015error}). In scale-free systems, which are likely to be all systems where a continuous symmetry is spontaneously broken,
we expect $\xi\sim L$, hence the method should work well as long as we manage to gather data on systems which 
are reasonably large. In generic, non-scale-free systems one should take care in determining the amount of static 
correlation before proceeding with the full fledged dynamical analysis.

One may ask whether the method is useful in generic biological data sets, for which we have no {\it a priori} reason to believe that either the Vicsek model, 
or the inertial spin model are correct. This is a very pertinent question and we are afraid that our answer may perhaps sound reasonable only to statistical physicists. 
The two models we have analyzed here are probably the {\it simplest} collective motion models with non-inertial and inertial dynamics. This is clear
by their mathematical structure: the differential spatial part is a Laplacian in both models, which is a very basic way to implement imitation, while the differential dynamical part is
 first-order 
in the Vicsek model and it is second-order in the inertial spin model, which is the minimal way to have the 
emergence of linear phase waves uncoupled to density waves. Many other models, different from both Vicsek and the ISM, can be envisaged, of course. However, 
we believe that the essential mathematical difference between inertial and non-inertial dynamics can hardly be represented by something radically different from what
we have described here.

\subsection*{Acknowledgements.}
This work was supported by IIT-Seed Artswarm, European Research Council Starting Grant 257126, and US Air Force Office of Scientific Research Grant FA95501010250 (through the University of Maryland). We thank William Bialek, Lorenzo Del Castello and Leonardo Parisi for discussions.

\appendix

%%%%%%%%%%%%%%%%%%%%%%%%%%%%
\section{Details of the numerical simulations}
%%%%%%%%%%%%%%%%%%%%%%%%%%%%

To simulate the Vicsek or ISM models means to integrate numerically
the corresponding equations.  In \cite{cavagna+al_15}, this was done with an Euler
method, but this has the disadvantage that it is not very stable when
the friction is low (i.e.\ when the inertial effects dominate).  Also,
the constraint $v_i(t)^2=v_0^2$  is not exactly enforced this way,
because the exact 
equations enforce it by requiring that $d{\bf v}_i/dt$ be perpendicular to
${\bf v}_i$, but this is not sufficient when using finite time differences.
For these reasons here we have resorted to an integration scheme used
in Brownian Dynamics, which allows for exact implementation of the
constraint via Lagrange multipliers and which in the underdamped
($\eta\to0$) case reduces to the velocity Verlet integrator used in
Molecular Dynamics, widely used due to its good energy conservation
properties and computational affordability \cite{Rapaport2004}.  The
only drawback is that the overdamped  (Vicsek) case with $m=0$ cannot
be integrated this way.  We thus have treated the Vicsek model
separately, with an Euler integrator, as in simulations of overdamped
Brownian motion in liquids \cite{Allen1987}.

\subsection{Integration of the ISM equations}
\label{sec:integr-ism-equat}

We start from the second order equation for the velocity, which we
rewrite as
\begin{equation}
  \frac{d^2{\bf v}_i}{dt^2} = \frac{v_0^2}{\chi} \left[
  {\bf F}_i(\{{\bf r}_j,{\bf v}_j\}) + {\bf F}_{v,i}  +
  {\bf f}_{c,i} \right], 
\label{eq:4}
\end{equation}
where the first two terms on the r.h.s.\ include the social
interaction, which is function of the positions
and velocities of the particles, and the random and viscous forces,
\begin{align}
  {\bf F}_i & = \frac{J}{v_0^2} \sum_j n_{ij} {\bf v}_j, \label{eq:6}\\
  {\bf F}_{v,i} &= - \frac{\eta}{v_0^2} \frac{d {\bf v}_i}{dt} + \frac{{\boldsymbol{\zeta}}_i}{v_0},
\end{align}
and the term ${\bf f}_{c,i}$ is the constraint force, given by the rest of the terms of
eq. \eqref{grand}, but which we compute differently in the
discretized equations, so that the constraint is exactly enforced.  To
obtain the discretized equations we integrate eq.~\ref{eq:4}
assuming ${\bf F}_i$ varies linearly in time in a small interval
$\Delta t$~\cite{allen_brownian_1980,Allen1987}.  The term
${\bf f}_{c,i}$ is disregarded at first, and later reintroduced  as explained below.  Defining
${\bf a}_i=d{\bf v}_i/dt$, ${\bf b}_i=d{\bf a}_i/dt$, one arrives
at
\begin{subequations}
  \begin{align}
    {\bf r}_i(t+\Delta t) ={}& {\bf r}_i(t) + \Delta t{\bf v}_i,\\
    {\bf v}_i(t+\Delta t) ={}& {\bf v}_i(t) + \Delta t c_1 {\bf a}_i(t) +
                               (\Delta t)^2 c_2 {\bf b}_i(t) + \notag \\
                             & (\Delta t)^2
                               c_2 \lambda_i(t) + \boldsymbol{\Xi}_v(t), \\
    {\bf a}_i(t+\Delta t) ={}& c_0 {\bf a}_i(t) + (c_1-c_2) \Delta t \left[
                               {\bf b}_i(t) + \lambda_i {\bf v}_i(t)\right] + \notag\\
                             &c_2 \Delta t \left[ {\bf b}_i(t+\Delta t) +
                               \mu_i {\bf v}_i(t+\Delta t) \right] +
                               \notag \\ & \boldsymbol{\Xi}_a(t),  \label{eq:5}
    \\
    {\bf b}_i(t+\Delta t) ={}& \frac{v_0^2}{\chi}
                               {\bf F}_i(\{{\bf r}_j(t+\Delta
                               t),{\bf v}_j(t+\Delta t)\}),
  \end{align}
\end{subequations}
where $\lambda_i$ and $\mu_i$ are related to the constraint (see
below) and the other constants result from the integration: $c_0$,
$c_1$, and $c_2$ are 
\begin{align}
  c_0 & = e^{-\eta v_0^2 \Delta t/\chi}, \\
  c_1 & = \frac{\chi}{v_0^2\eta \Delta t} \big(1-c_0\big), \\
  c_2 & = \frac{\chi}{v_0^2\eta \Delta t} \big(1-c_1\big),
\end{align}
and $\boldsymbol{\Xi}_v$ and $\boldsymbol{\Xi}_a$ are random variables related to the random force.
They are independent for each axis, and each pair
of components is drawn from a bivariate Gaussian distribution with
zero first moments and second moments given by
\begin{align}
  \langle \Xi_v^2 \rangle &= \frac{Tv\chi}{v_0^2\eta}\left(
                            2\frac{\eta v_0^2\Delta t}{\chi} -3 + 4
                            e^{-\eta v_0^2\Delta t/\chi} -e^{-2\eta
                            v_0^2 \Delta t/\chi} \right), \notag\\
  \langle \Xi_a^2 \rangle &= \frac{Tv_0^2}{\chi} \left( 1- e^{-2\eta
                            v_0^2\Delta t/\chi} \right), \\
  \langle \Xi_v\Xi_a\rangle &= \frac{T}{\eta}\left(1 - e^{-\eta v_0^2
                              \Delta t/\chi} \right).\notag
\end{align}
This scheme has the advantage that it reduces to the velocity Verlet
integrator for Molecular Dynamics~\cite{Allen1987,swope_computer_1982-1}
in the underdamped $\eta\to0$ limit, which is known to stably
reproduce the energy conservation property of Newton's equations (in
our case applying to the conservation of the Hamiltonian in the case
$\eta=0$ for metric interactions on a fixed arbitrary lattice).

The constraint is enforced as in the RATTLE
algorithm~\cite{andersen_rattle:_1983}, only that since the
constraints on each particle are independent, the Lagrange
multipliers can be found analytically and there is no need of an
iterative procedure.  Imposing $v_i^2(t+\Delta t)=v_0^2$ and
${\bf v}_i(t+\Delta t) \cdot {\bf a}_i(t+\Delta t)=0$ one obtains
\begin{align*}
  \lambda_i ={}& \frac{w_+-1}{(\Delta t)^2 c_2},\\
  \mu_i= {}&- \frac{{\bf v}_i(t+\Delta t) \cdot {\bf a'}_i(t+\Delta
  t)}{c_2 v_0^2 \Delta t},
\end{align*}
where $w_+$ is the positive root of
\begin{gather}
  \label{eq:3}
  v_0^2 w^2 + 2{\bf v}_i(t)\cdot\Delta{\bf v}_i w +\Delta v_i^2=v_0^2,
  \notag \\
  \Delta{\bf v}_i = c_1\Delta t{\bf a}_i(t) + c_2(\Delta t)^2{\bf b}_i(t),
\end{gather}
and ${\bf a'}_i(t+\Delta t)$ is equal to ${\bf a}_i(t+\Delta t)$ as
given by eq.~(\ref{eq:5}) but without the term proportional to
$\mu_i$.

Each step is performed in two stages, as in the velocity Verlet
scheme~\cite{Allen1987}:  First the random variables are drawn,
${\bf r}_i$ is updated, 
${\bf a}_i$ is partially updated using only the terms that depend on
quantities evaluated at $t$; the ${\bf v}_i$ are updated, and the
constraint terms computed and applied.  Then the force at the new
positions and velocities is computed, and finally the update of ${\bf a}_i$ is
completed.

\subsection{Integration of the Vicsek equations}
\label{sec:integr-vics-model}

The Vicsek model (eqs.~\ref{vic1}, \ref{vic2}) is the overdamped
($\chi/\eta^2\to0$) limit of the ISM, but although the above scheme works
very well for $\eta\to0$, it is not suitable for the overdamped case,
which is equivalent to setting $\chi=0$.  We thus use a simple Euler
integration \cite{Allen1987}, derived by integrating eq.~\ref{vic1}
over $\Delta t$ assuming ${\bf F}_i(\{{\bf r}_i,{\bf v}_i\})$
constant, and enforcing the constraint as before.  Setting $\eta=1$
(which amounts to a rescaling of time), this results in
\begin{align}
  {\bf r}_i(t+\Delta t) ={}& {\bf r}_i(t) + \Delta t{\bf v}_i,\\
  {\bf v}_i(t+\Delta t) ={}& v_0^2 \left[ \Delta t {\bf F}_i(t) +
                             \boldsymbol{\Xi}_i, \right]  + w_i
                             {\bf v}_i(t),
\end{align}
where $w_i$ is the smallest solution of
\begin{gather}
  \label{eq:3bis}
  v_0^2 w^2 + 2{\bf v}_i(t)\cdot\Delta{\bf v}_i w +\Delta v_i^2=v_0^2,
  \notag \\
  \Delta{\bf v}_i = v_0^2\Delta t{\bf F}_i(t) + \boldsymbol{\Xi}_i,
\end{gather}
and $\boldsymbol{\Xi}$ are  Gaussian random variables, independent for each
axis, of zero mean and variance
\begin{equation}
  \label{eq:7}
  \sigma^2_\Xi = \frac{2 T}{v_0^2} \Delta t.
\end{equation}

\subsection{Parameters and runs}
\label{sec:runs}

We performed numerical simulations on both Vicsek and ISM models in $d=3$ on a cube with periodic boundary conditions, for systems of different sizes: $N = 512, 724, 1024, 1448, 2048$. In all cases the density was fixed, $\rho=N/L^3=0.147$, corresponding to a mean interparticle distance, $a\sim 1$. 
For both models we choose the following parameters: temperature $T=1$, friction $\eta=1$, strength of the interaction $J=1$. These parameters 
correspond to polarization $\Phi\sim 0.9$.

In the ISM model we fixed $\chi=5$, so that $\gamma=\eta/2\chi=0.1$, $c_s=\sqrt{Jn_ca^2/\chi}=1.79$ and $k_0=\gamma/c_s=0.056$. This choice of the parameters guarantees that the ISM simulations are in the underdamped regime, because $k_0=0.056\ll 1/L$ for all the analyzed systems: $2\pi/L \in[0.26: 0.41]$, where the lower bound corresponds to the biggest system ($N=2048$, $L=24$) and the upper bound to the smallest system ($N=512$, $L=15$). In terms of the dispersion relation depicted in Fig.5, we can say that in all our systems the physical momentum $k$
is always much larger than the edge of overdamping, $k_0$.

We run simulations with both topological and  metric interaction.
In the topological case the number of interacting neighbours is $n_c=16$; in the metric case the interaction range is $r_c=2.95$, such that, on average, each particle has $n_c\sim16$ interacting neighbors and a fair comparison between topological and metric interaction is possible.  We find no significant difference of the scaling laws and critical exponents between the two cases (shown in the figures are topological results).
 
Systems were initialized in order to have randomly distributed particles with all the velocities directed along the $x$-axis, ${\bf v} = v_0 \; (1, 0, 0)$ and $v_0=0.1$.
Vicsek simulations have a total duration $t_\mathrm{tot}= 6\times 10^5$ time steps, while ISM simulation have a total duration $t_\mathrm{tot}=10^6$ time steps; in both cases we saved the particles  position and velocity at intervals of $10^2$ time steps, we analyzed $6$ samples of the duration of $10^4$ time steps for each simulation and we averaged the correlation functions on the different samples in order to reduce fluctuations.

%%%%%%%%%%%%%%%%%%%%%%%%%%%%%%%%%%%%%%%%%%%%%%%%%%%%%%%%%%%%%%%

%%%%%%%%%%%%%%%%%%%%%%%%%%%%%%%%%%%
\section{Green functions method}
%%%%%%%%%%%%%%%%%%%%%%%%%%%%%%%%%%%

To fix ideas we will describe the Green functions method \cite{lanczos1961linear} by using as an example the Vicsek model in its
continuous limit form,
\begin{equation}
\left(\eta \frac{\partial }{\partial t} - J n_c a^2 \, \nabla^2 \right)\varphi({\bf x}, t) =\zeta({\bf x}, t) 
\end{equation}
To solve this  linear stochastic equation it is convenient to first find the solution of the following Green equation,
\beq
\left(\eta \frac{\partial }{\partial t} - J n_c a^2 \, \nabla^2 \right) G({\bf x}, t) = \delta^{(3)}({\bf x})\delta(t) 
\label{green}
\eeq
where $G({\bf x}, t)$ is the Green equation (or dynamical propagator) associated to the original dynamical equation. Once we know the Green function, 
we can write the general solution of the original equation (up to a solution of the homogenous problem) as, 
\beq
\varphi({\bf x}, t)= \int d{\bf x}' dt' \ G({\bf x} - {\bf x}', t - t') \zeta({\bf x}', t')
\eeq
It is convenient at this point to switch to a Fourier representation in terms of momentum $\bf k$ and frequency $\omega$,
\begin{subequations}
\begin{align}
G({\bf x}, t)= \frac{1}{(2 \pi)^4}\int d{\bf k} \, d \omega \ e^{i \left({\bf k}\cdot {\bf x} - \omega t \right)} \, G({\bf k}, \omega) \\
\varphi({\bf x}, t)= \frac{1}{(2 \pi)^4}\int d{\bf k} \, d \omega \ e^{i \left({\bf k}\cdot {\bf x} - \omega t \right)} \, \varphi({\bf k}, \omega) 
\end{align}
\end{subequations}
so that the previous equations become polynomial,
\beq
(i \eta \omega+ J n_c a^2 k^2) G({\bf k}, \omega) = 1 \\ 
\eeq
In this way one obtains a simple algebraic expression for the dynamical Green function,
\begin{equation}
G({\bf k}, \omega) = \frac{1}{ i \eta \omega + J n_c a^2 k^2}
\label{grongobis}
\end{equation}
Clearly, the Green function $G({\bf k}, \omega)$ contains all the relevant information to infer the dispersion relation, and thus the
full dynamical equation ruling the system. The most direct way to access $G({\bf k}, \omega)$ is to compute the correlation of
the field $\varphi({\bf k}, \omega)$: the solution of the dynamical equation in Fourier space is,
\beq
\varphi({\bf k}, \omega) = G({\bf k}, \omega)\, \zeta({\bf k}, \omega) 
\eeq
so that if we now multiply two fields and average over the noise we get,
\bea
C({\bf k}, \omega)\equiv & \, \langle \varphi({\bf k}, \omega) \varphi(-{\bf k}, -\omega) \rangle \quad \quad\quad \quad
\\
\quad \quad = &2 \eta T \, a^3G({\bf k}, \omega) \, G(-{\bf k}, -\omega) 
\eea
By doing the  Fourier integral in the frequency we finally obtain the spatio-temporal correlation function in Fourier space, $C({\bf k},t)$, 
\beq
C({\bf k}, t) =\frac{2 \eta T\,a^3}{2 \pi} \int d\omega \,  \, e^{i \omega t} \ G({\bf k}, \omega) \, G(-{\bf k}, -\omega) 
\label{zutbis}
\eeq
This integral is solved in general by Cauchy residue method, so that the poles of the Green function acquire particular importance. It is 
for this reason that one needs to write the so-called {\it dispersion relation} associated to the original dynamical equation,
\beq 
i \eta \omega+ J n_c a^2 k^2=0\ .
\eeq

%%%%%%%%%%%%%%%%%%%%%%%%%%%%%%%%%%%%%%%%%%%%%%%%%%%%%%%%%%%%%%%

\bibliographystyle{apsrev4-1}
\bibliography{irene_byhand}

\end{document}